\pdfoutput=1

\documentclass[11pt,twoside,a4paper,cmspaper,final,collab]{cms-tdr}
\def\svnVersion{48f131f-D}\def\svnDate{2020/04/18}\def\cmsCernNoTag{CERN-EP-2020-007}\def\cmsCernDate{\today}\def\cmsMessage{"Published in the Journal of High Energy Physics as \href{http://dx.doi.org/10.1007/JHEP06(2020)076}{\doi{10.1007/JHEP06(2020)076}.}"}

\begin{document}\cmsNoteHeader{SMP-18-007}

\hyphenation{had-ron-i-za-tion}
\hyphenation{cal-or-i-me-ter}
\hyphenation{de-vices}
\RCS$HeadURL: svn+ssh://svn.cern.ch/reps/tdr2/papers/SMP-18-099/trunk/SMP-18-099.tex $
\RCS$Id: SMP-18-099.tex 465952 2018-06-25 12:42:44Z glandsbe $
\newlength\cmsFigWidth
\newlength\cmsTabSkip\setlength{\cmsTabSkip}{1.7ex}
\ifthenelse{\boolean{cms@external}}{\setlength\cmsFigWidth{0.85\columnwidth}}{\setlength\cmsFigWidth{0.4\textwidth}}
\newcommand{\mll}{m_{\ell\ell}}
\newcommand{\mjj}{m_\mathrm{jj}}
\newcommand{\mzg}{m_{\PZ\gamma}}
\newcommand{\etajj}{\Delta \eta_{\mathrm{jj}}}
\newcommand{\gbarrel}{\gamma_{\text{barrel}}}
\newcommand{\gendcap}{\gamma_{\text{endcap}}}
\newcommand{\dphizgjj}{\Delta \phi_{\PZ\gamma, \mathrm{jj}}}

\cmsNoteHeader{SMP-18-099}
\title{Measurement of the cross section for electroweak production of a {\PZ} boson, a photon and two jets in proton-proton collisions at $\sqrt{s} = 13\TeV$ and constraints on anomalous quartic couplings}

\date{\today}

\abstract{
A measurement is presented of the cross section for electroweak production of a {\PZ} boson and a photon in association with two jets ($\PZ\gamma$jj) in proton-proton collisions. The {\PZ} boson candidates are selected through their decay into a pair of electrons or muons. The process of interest, electroweak $\PZ\gamma$jj production, is isolated by selecting events with a large dijet mass and a large pseudorapidity gap between the two jets. The measurement is based on data collected at the CMS experiment at $\sqrt{s} = 13\TeV$, corresponding to an integrated luminosity of 35.9\fbinv. The observed significance of the signal is 3.9 standard deviations, where a significance of 5.2 standard deviations is expected in the standard model. These results are combined with published results by CMS at $\sqrt{s} = 8\TeV$, which leads to observed and expected respective significances of 4.7 and 5.5 standard deviations. From the 13\TeV data, a value is obtained for the signal strength of electroweak $\PZ\gamma$jj production and bounds are given on quartic vector boson interactions in the framework of dimension-eight effective field theory operators.}

\hypersetup{
pdfauthor={CMS Collaboration},
pdftitle={Measurement of the cross section for electroweak production of a Z boson, a photon and two jets in proton-proton collisions at sqrt(s) = 13 TeV and constraints on anomalous quartic couplings},
pdfsubject={CMS},
pdfkeywords={CMS, physics, vector boson scattering}}

\maketitle
\section{Introduction}
The standard model (SM) is well tested and continues to be a reliable model of nature, bolstered by the discovery and measurement of the properties of the Higgs boson at the CERN LHC~\cite{Aad:2012tfa,Chatrchyan:2012ufa,Chatrchyan:2012ufa_long,higgs_measure_2016,higgs_measure_2019}. However, a search for incontrovertible evidence of new physics, and a lack of understanding of how all the forces unify motivates further study of the EW sector. For example, novel processes, such as vector boson scattering (VBS), probe unexplored aspects of the nonabelian nature of gauge interactions. The VBS processes are pure electroweak interactions where vector bosons are radiated from the initial state quarks and directly interact via scattering to produce a final state of two scattered vector bosons and two jets from the quarks. Many beyond-the-SM (BSM) models alter the couplings of vector bosons, and such effects can be parametrized through effective field theories~\cite{paper_aqgc}. The VBS topology is sensitive to quartic gauge couplings (QGCs) in the SM and to possible anomalous QGCs (aQGCs)~\cite{Eboli:2006wa}. Among all VBS categories, only VBS ZZ and VBS $\PZ\gamma$ are sensitive to pure neutral aQGCs. The VBS $\PZ\gamma$ has a larger cross section and tight limits are set in this paper.

The EW production of {\PW} boson pairs of the same charge was reported by the CMS and ATLAS experiments at $\sqrt{s} = 13\TeV$ at respective significances of 5.5 and 6.9 standard deviations~\cite{atlas:ssww,Sirunyan:2017ret}. The EW production of {\PW}{\PZ} bosons was also observed by ATLAS at 13\TeV at a significance of 5.3 standard deviations~\cite{atlas_WZ}. Measurements of the EW production cross section of a {\PZ} boson and a photon were reported by CMS and ATLAS, based on earlier data collected at 8\TeV, corresponding to respective integrated luminosities of 19.7 and 20.2\fbinv~\cite{Aaboud:2017pds,Khachatryan:2017jub}. The observed significances of these measurements were respectively 3.0 and 2.0 standard deviations for CMS and ATLAS, where respective significances of 2.1 and 1.8 standard deviations were expected based on the SM; limits were also reported on the aQGCs. Recently, measurements of the EW production of $\PZ\gamma$ bosons were updated by ATLAS based on data collected at 13\TeV at a significance of 4.1 standard deviations~\cite{atlas_zg_13}.

We present a study of EW production of $\PZ\gamma$jj that includes a measurement of the production cross section and limits on aQGCs at 13\TeV. The data correspond to an integrated luminosity of $35.9\pm 0.9\fbinv$ of proton-proton ($\Pp\Pp$) collisions collected using the CMS detector in 2016.
Candidate events are selected to contain: (i) two identified leptons (electrons or muons) that come from either direct {\PZ} boson decay or from indirect {\PZ} boson decay through the {\PZ} $\to \tau \tau$ chain; (ii) one identified photon; (iii) two jets with a large separation in pseudorapidity ($\eta$); and (iv) a large dijet mass. The jet selection reduces the contribution from the non-VBS production of $\PZ\gamma$, making this signature an ideal topology for VBS studies.

\begin{figure}[h]
\centering
      \includegraphics[width=0.32\textwidth]{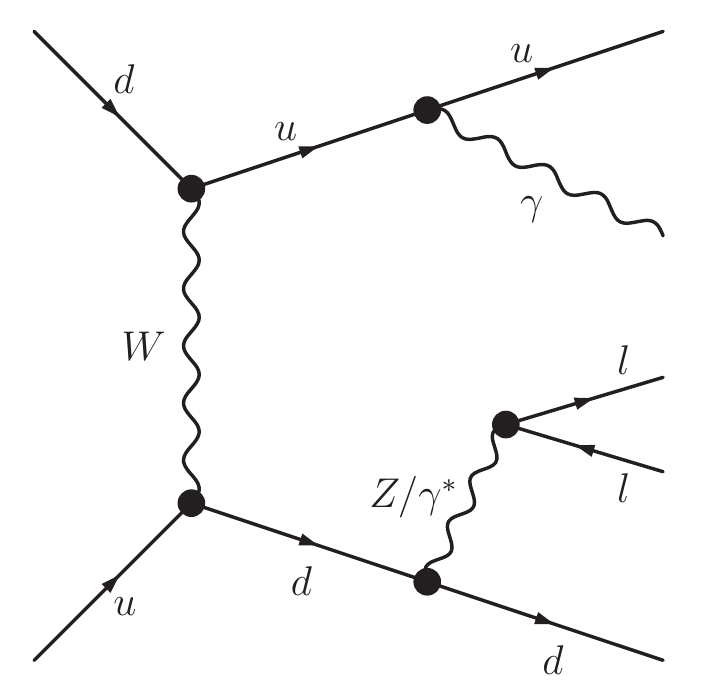}
      \includegraphics[width=0.32\textwidth]{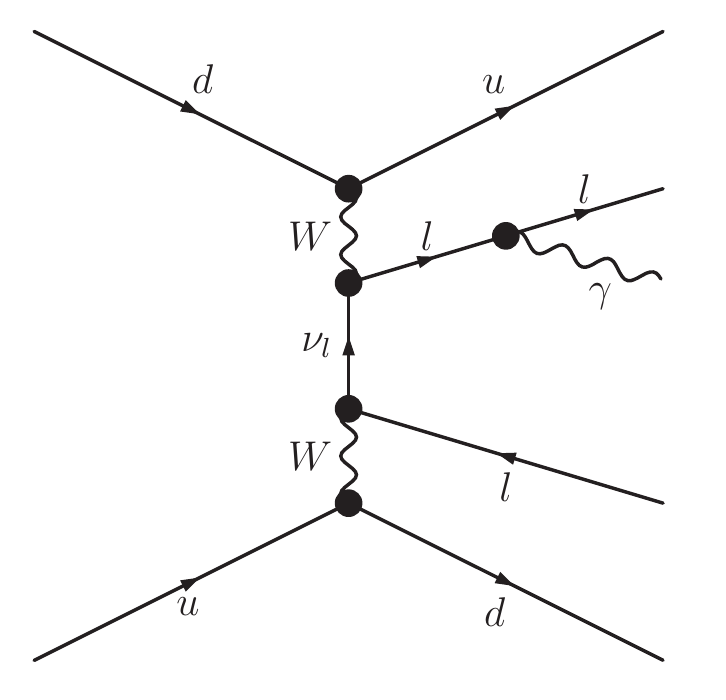}
      \includegraphics[width=0.32\textwidth]{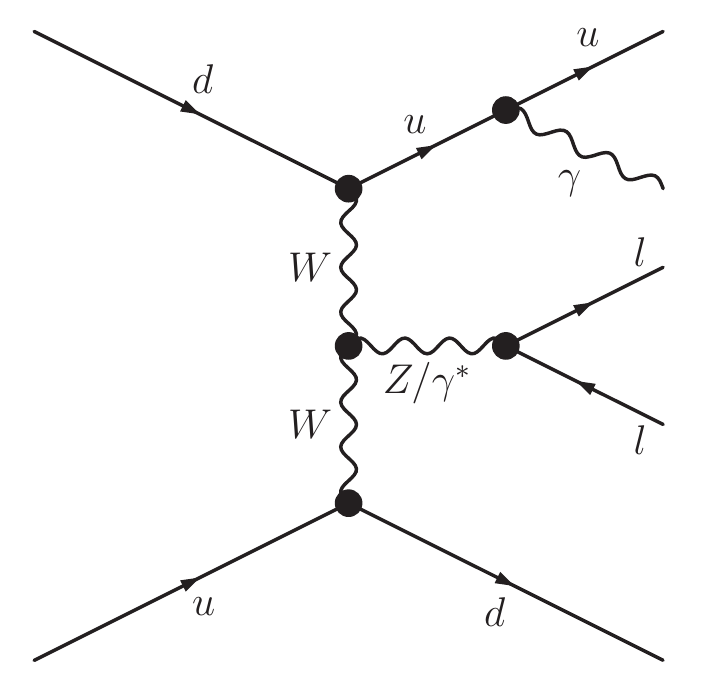}\\
      \includegraphics[width=0.32\textwidth]{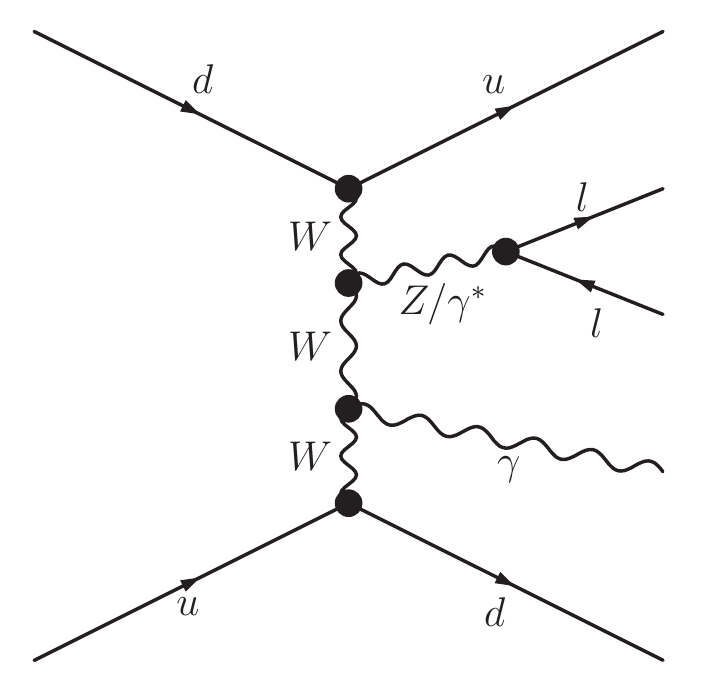}
      \includegraphics[width=0.32\textwidth]{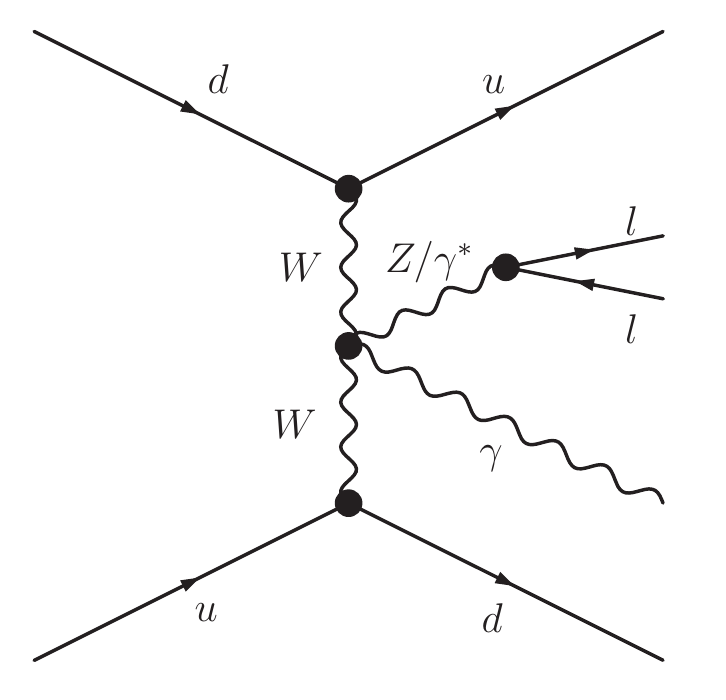}
      \includegraphics[width=0.32\textwidth]{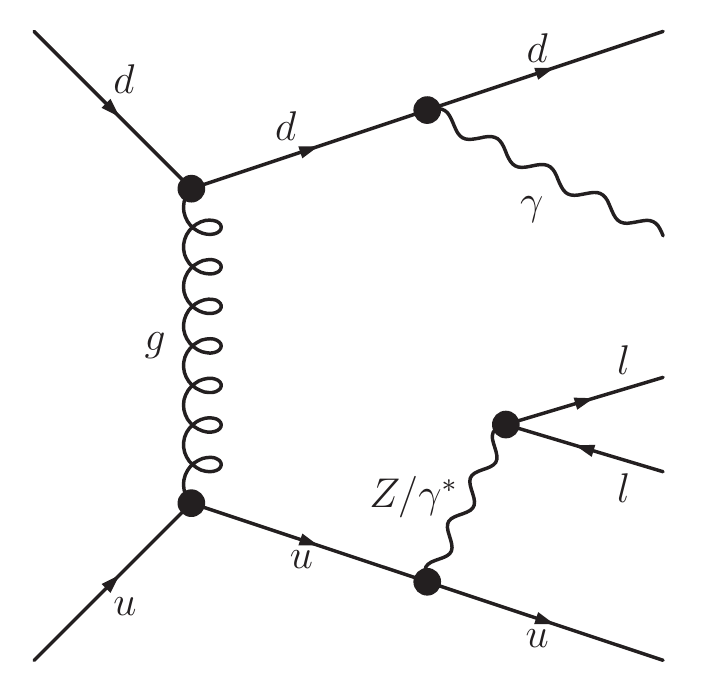}
\caption{Representative Feynman diagrams for $\PZ\gamma$jj production. The diagrams except (lower right) reflect EW origin: (upper left) bremsstrahlung, (upper center) multiperipheral, (upper right) VBF with TGCs, (lower left) VBS via {\PW} boson, (lower center) VBS with QGCs, while (lower right) is a QCD-induced diagram.}
\label{fig:za_feynman}
\end{figure}

Figure~\ref{fig:za_feynman} shows representative Feynman diagrams, including (upper left)~bremsstrahlung, (upper center)~multiperipheral (or nonresonant) production, (upper right)~vector boson fusion with trilinear gauge boson couplings (TGCs), (lower left)~VBS via a {\PW} boson, (lower center)~VBS via QGC, and (lower right)~quantum chromodynamics (QCD) induced production of $\PZ\gamma$. The VBS processes are particularly interesting because they involve QGCs (\eg $\PW\PW\PZ\gamma$). It is not possible, however, to isolate the QGC diagrams from the other contributions that are topologically similar, such as VBS via W boson diagrams. The EW production mechanisms of order $\alpha^5$ at lowest ``tree'' level are regarded as signal, and the QCD-induced production mechanisms of order $\alpha^3\alpS^2$ at ``tree'' level are regarded as background, where $\alpha$ and $\alpS$ are the respective electromagnetic and strong couplings.

\section{The CMS detector}
The central feature of the CMS~\cite{Chatrchyan:2008zzk} apparatus is a superconducting solenoid of 6\unit{m} internal diameter, providing a magnetic field of 3.8\unit{T}. A silicon pixel and strip tracker, a lead tungstate crystal electromagnetic calorimeter (ECAL), and a brass and scintillator hadron calorimeter (HCAL), each composed of a barrel and two endcap sections reside within the solenoid volume. Forward calorimeters extend the coverage provided by the barrel and endcap detectors up to pseudorapidities of $\abs{\eta}=5$. Muons are measured in gas-ionization detectors embedded in the steel flux-return yoke outside the solenoid.

Events of interest are selected using a two-level trigger system~\cite{trigger_system}. The first level (L1), composed of specialized hardware processors, uses information from the calorimeters and muon detectors to select events of interest with a maximum rate of 100\unit{kHz}. A high-level trigger processor farm decreases this rate to 1\unit{kHz} before storage. A more detailed description of the CMS detector, together with a definition of the coordinate system and kinematic variables, can be found in Ref.~\cite{Chatrchyan:2008zzk}.

\section{Signal and background simulation}

The signal and the main background (QCD-induced $\PZ\gamma$jj) processes are simulated using the respective \MGvATNLO~2.3.3 and 2.6.0~\cite{MGatNLO} Monte Carlo (MC) generators. The EW $\PZ\gamma$jj signal is simulated at leading order (LO) in QCD, and the QCD-induced $\PZ\gamma$jj process simulated at up to one jet in the matrix element calculations at next-to-leading-order (NLO) in QCD, using the FxFx jet merging scheme~\cite{Frederix:2012ps}. The magnitude of the interference is 4--8\% depending on $\mjj$ and is described in the section on systematic uncertainties. Other background contributions arise from two general classes. The VV backgrounds include QCD-induced $\PW\gamma$jj production simulated at NLO using \MGvATNLO~2.6.0 and diboson processes {\PW}{\PW}/{\PW}{\PZ}/{\PZ}{\PZ} simulated using \PYTHIA 8.212~\cite{Sjostrand:2014zea}. Top backgrounds include single top quark production simulated at NLO using \POWHEG 2.0~\cite{POWHEG,POWHEG1,POWHEG2,POWHEG3} and $\ttbar\gamma$ production simulated at NLO with \MGvATNLO~2.2.2 using the FxFx jet matching scheme.

{\tolerance=10
The simulation of the inclusion of a aQGC is performed using \MGvATNLO~2.2.2 at LO. The matrix element reweighting feature in \MGvATNLO that provides each event with additional weights corresponding to different theoretical hypotheses, \eg, a different model or a different choice of parameters, is used to extract the size of the coefficients of any anomalous coupling operators probed in the analysis~\cite{mg5:reweight}.\par}

The \PYTHIA 8 generator package using the CUETP8M1 tune is used for parton showering, hadronization, and simulating the underlying event~\cite{Skands:2014pea,Khachatryan:2015pea}. The NNPDF~3.0~\cite{nnpdf} parton distribution functions (PDFs) are used in these studies, and the CMS detector response in simulated events is modeled using the \GEANTfour package~\cite{geant4, geant4_2}. A tag-and-probe procedure~\cite{tagandprobe} is used to measure factors to correct for data-to-simulation differences in trigger, reconstruction, and selection efficiencies. The simulated events include additional $\Pp\Pp$ interactions in the same and neighboring bunch crossings, referred to as pileup (PU). Simulated events are weighted so the PU distribution matches the one from data, with an average PU of $\approx$23 interactions per bunch crossing.

\section{Object reconstruction and event selection}
\subsection{Objects reconstruction}
\label{sec:event rec}
A particle-flow (PF) algorithm~\cite{Sirunyan:2017ulk} is used to reconstruct particles in the event. It combines all subdetector information to reconstruct individual objects and identify them as charged or neutral hadrons, photons, or leptons (PF candidates).

The reconstructed vertex with largest value in summed object $\pt^2$ defines the primary $\Pp\Pp$ interaction vertex~\cite{pv_1} (where \pt is the transverse momentum). The objects can also refer to jets clustered using a jet finding algorithm~\cite{antikt,Cacciari:fastjet1} and hadrons assigned to the vertex as inputs. The associated imbalance in transverse momentum in the event ($\ptmiss$) is the magnitude of the vector \pt sum of these jets.

Electrons are reconstructed within $\abs{\eta}<2.5$ for $\pt>25\GeV$. This involves combining the information from clusters of energy deposited in the ECAL and the trajectories fitted in the tracker~\cite{Khachatryan:2015hwa}. The energies of electrons are evaluated from a combination of the electron momentum at the primary interaction vertex determined in the tracker, the energy in the corresponding ECAL cluster, and the energy sum of all bremsstrahlung photons spatially compatible with originating from the electron track. To reduce electron misidentification, electron candidates are required to pass additional identification criteria based on the relative amount of energy deposited in the HCAL, a match of the trajectory in the inner tracker with that in the supercluster~\cite{Khachatryan:2015hwa} of the ECAL, the number of missing hits in the inner tracker, the consistency between the track and the primary vertex, and $\sigma_{\mathrm{i}\eta\mathrm{i}\eta}$, a parameter that quantifies the spread in $\eta$ of the electromagnetic shower in the ECAL, as discussed in Section~\ref{sec:bkg_estimation}. Electron candidates identified as originating from photon conversions are rejected~\cite{Khachatryan:2015hwa,electron_7tev}. Different working points are defined according to their efficiency. The ``medium'' working point is used to reconstruct electrons in the final state, and a much less restrictive working point, referred to as ``veto'', is used to reconstruct electrons for vetoing events that contain additional leptons. The medium categories have efficiencies of $\approx$80\% for acceptance of signal and $\approx$99\% for background rejection that change the respective values to 95 and 96\% for the veto working point.

Muons are reconstructed from information in the muon system and the inner tracker at $\abs{\eta}<2.4$ and $\pt>20\GeV$~\cite{muon_13tev}. The energies of muons are obtained from the curvature of the corresponding tracks. Muon candidates must satisfy identification criteria based on the number of hits in the muon system and the inner tracker, the quality of the combined fit to a track, the number of matched muon-detector planes, and the consistency between the track and the primary vertex. Different working points are defined according to their efficiency. A highly restrictive working point is used to reconstruct muons in the final state, and a far less restrictive working point, referred to as ``minimal'', is used to reconstruct muons for vetoing events with additional leptons.

Additional cutoffs on relative isolation variables are applied for both electrons and muons. These are defined relative to their \pt values by summing the \pt of the charged hadrons and neutral particles in geometrical cones $\Delta R = \sqrt{\smash[b]{(\Delta\eta)^2+(\Delta\phi)^2}} = 0.3$ or 0.4, respectively, about the electrons and muons trajectories:
\begin{linenomath}
\begin{equation*}\label{iso}
\text{Iso} = \left(\sum\pt^{\text{charged}} + \text{MAX}\left[0, \sum\pt^{\text{neutral}} + \sum\pt^{\gamma} - \pt^{\mathrm{PU}}\right]\right)/\pt,
\end{equation*}
\end{linenomath}
where $\sum\pt^{\text{charged}}$ is the scalar \pt sum of charged hadrons originating from the primary vertex; and $\sum\pt^{\text{neutral}}$ and $\sum\pt^{\gamma}$ are the respective scalar \pt sums of neutral hadrons and photons. The contribution from PU in the isolation cone, \ie, $\pt^{\mathrm{PU}}$, is subtracted using the \FASTJET technique~\cite{Cacciari:fastjet1}. For electrons, $\pt^{\mathrm{PU}}$ is evaluated using the ``jet area'' method described in Ref.~\cite{jetarea_method}. For muons, $\pt^{\mathrm{PU}}$ is assumed to be half of the scalar \pt sum deposited in the isolation cone by charged particles not associated with the primary vertex. The factor of 0.5 corresponds approximately to the ratio of neutral to charged hadrons produced in the hadronization of PU interactions. Electrons are considered isolated when the respective working points for medium and veto are set to $\text{Iso}<0.0695$ or $<0.175$ in the barrel, or $\text{Iso}<0.0821$ or $<$0.159 in the endcap detector regions. Muons are considered isolated when $\text{Iso}<0.15$ or $<$0.25 for the respective highly restrictive and minimal working points.

Photon reconstruction and selections are similar to those for electrons, and performed in the region of $\abs{\eta}<2.5$~\cite{photon_8tev} and $\pt>20\GeV$, excluding the ECAL transition region of $1.444<\abs{\eta}<1.566$. The energies of photons are obtained from the ECAL measurements. Photons located in the barrel region, $0<\abs{\eta}<1.444$ and the endcap ECAL region, $1.566<\abs{\eta}<2.5$, will be referred to as $\gbarrel$ and $\gendcap$, respectively. To minimize photon misidentification, photon candidates are required to pass an electron veto, and satisfy criteria based on the distribution of electromagnetic energy in the ECAL and in the HCAL, and on the isolation variables constructed from the kinematic inputs of the charged and neutral hadrons, and other photon candidates present near the photon of interest. The medium working point is used to reconstruct prompt photons (\ie, not from hadron decays) in the final state, and the minimal working point used to reconstruct nonprompt photons that are mainly products of neutral pion decay~\cite{photon_8tev}.

Jets are reconstructed using PF objects and the anti-\kt jet clustering algorithm~\cite{antikt} with a distance parameter of 0.4. The energies of charged hadrons are determined from a combination of their momenta measured in the tracker and the matching of ECAL and HCAL energy deposits, corrected for the response of the calorimeters to hadronic showers. The energy of neutral hadrons is obtained from the corresponding corrected ECAL and HCAL energies. To reduce the contamination from PU, charged PF candidates in the tracker acceptance of $\abs{\eta}<2.4$ are excluded from jet clustering when they are associated with PU vertices~\cite{Sirunyan:2017ulk}. The contribution from neutral PU particles to the jet energy is corrected based on the projected area of the jet on the front face of the calorimeter. Jets are required to have $\pt>30\GeV$ and $\abs{\eta}<4.7$. A jet energy correction, similar to the one developed for 8\TeV collisions~\cite{jer}, is obtained from a dedicated set of studies performed on both data and MC events (typically involving dijet, photon+jet, {\PZ}+jet and multijet production). Other residual corrections are applied to the data as functions of $\pt$ and $\eta$ to correct for the small differences between data and simulation. Additional quality criteria are applied to jet candidates to remove spurious jet-like features originating from isolated noise patterns in the calorimeters or in the tracker.

\subsection{Event selection}
\label{sec:event_selection}
Collisions are selected in data using triggers that require the presence of one or two electrons or two muons. The lepton with highest \pt is referred to as the leading lepton and denoted $\ell 1$, and the lepton with second-highest \pt is referred to as the subleading lepton and denoted $\ell 2$. The \pt thresholds for $\ell 1$ and $\ell 2$ in the dilepton triggers are 23 and 12 for electrons, and 17 and 8\GeV for muons. For the single-electron trigger, the \pt threshold is 25\GeV. Partial mistiming of signals in the forward region of the ECAL endcap detectors ($2.5<\abs{\eta}<3.0$) resulted in L1 triggers being wrongly associated with the previous bunch crossing. Since rules for L1 triggers forbid two consecutive bunch crossings to fire, events with mistimed signals can self veto, which can lead to a significant decrease in L1 trigger efficiency. The loss of efficiency for EW $\PZ\gamma$jj events associated with the mistiming is $\approx$8\% for invariant mass of two jets $\mjj>500\GeV$, and increases to $\approx$15\% for $\mjj>2\TeV$. This effect is not taken into account in the simulation, and a correction is therefore applied as a function of jet \pt and $\eta$ using an unbiased data sample with correct timing. The correction is implemented through a factor that represents the probability of the event not having mistimed signals.

A selected event is required to have two oppositely charged same-flavor leptons for the reconstruction of a \PZ boson, \ie, either a pair of electrons or a pair of muons. All leptons used for the \PZ boson reconstruction must pass the more stringent identification and isolation requirements described in Section~\ref{sec:event rec}. The invariant mass of the dilepton system ($\mll$) must satisfy $70<\mll<110\GeV$. Events with a third lepton satisfying weaker identification criteria are rejected to reduce background from {\PW}{\PZ} and {\PZ}{\PZ} events.

Selected events are also required to contain at least one photon satisfying the identification criteria discussed in Section~\ref{sec:event rec}. The photon with largest \pt in the event is used when more than one passes the identification criteria. The $\Delta R$ between selected photons and selected leptons is required to be larger than 0.7. The invariant mass of the dilepton-photon system ($\mzg$) must satisfy $\mzg>100\GeV$ to reduce the contribution from final-state radiation in {\PZ} boson decays. Furthermore, the event must have at least two jets. The jet with largest \pt is called the leading jet, referred to as j1, and the jet with second-largest \pt is called the subleading jet, referred to as j2. Our selection of jets, leptons, and photons is referred to as the ``common'' selection.

A low-$\mjj$ control region, where the EW signal is negligible compared to QCD-induced $\PZ\gamma$jj production, is defined by the common selection and the requirement $150<\mjj<400\GeV$.

To exploit the unique signature of the VBS process, the leading plus subleading jet system is required to have an invariant mass greater than 500\GeV and an $\eta$ separation between the jets of $\etajj = \abs{\eta_{\mathrm{j1}} - \eta_{\mathrm{j2}}}>2.5$. The Zeppenfeld variable~\cite{zeppenfeld} $\eta^* = \abs{\eta_{\PZ\gamma} - (\eta_{\mathrm{j1}} + \eta_{\mathrm{j2}})/2}$, where $\eta_{\PZ\gamma}$ is the $\eta$ of the $\PZ\gamma$ system, is required to be $<2.4$. The expected recoil between the $\PZ\gamma$ and the dijet system, the variable $\dphizgjj$, the magnitude of the difference in azimuthal angle between the $\PZ\gamma$ and the dijet system, is required to be larger than 1.9. The constraints for $\eta^*$ and $\dphizgjj$ are optimized through simulation. This selection defines the EW signal region.

The cross sections for EW $\PZ\gamma$jj and EW+QCD $\PZ\gamma$jj production are measured in a fiducial region designed to approximate the acceptance of the CMS detector and the signal selection requirements based on the particle-level objects: (i) electrons and muons are required to be prompt, and those from $\tau$ lepton decays are excluded; (ii) the momenta of prompt photons with $\Delta R_{\ell\gamma}<0.1$ are added to the lepton momenta to correct for final-state photon radiation, referred to as ``dressing''; (iii) nonprompt photons are excluded; and (iv) VBS-like selections, \ie, $\mjj>500\GeV$ and $\etajj>2.5$ are required. Additional selections on electrons, muons, photons, and jets are the same as defined in the common selection.

The aQGC search is performed in a region similar to the fiducial region, but with the additional requirement of $\pt^{\gamma}>100\GeV$.

A summary of all the selection criteria for the various regions is shown in Table~\ref{tab:selections}.

\begin{table}[htbp]
\centering
\topcaption{Summary of the five sets of event-selection criteria used to define events in the common selection, control region selection, EW signal extraction, the fiducial cross section, and the search for an aQGC contribution.}
\begin{tabular}{lc}
\hline
 Common selection & $\pt^{\ell1,\ell2}>25\GeV$, $\abs{\eta^{\ell1,\ell2}}<2.5$ for electron channel\\
 		  & $\pt^{\ell1,\ell2}>20\GeV$, $\abs{\eta^{\ell1,\ell2}}<2.4$ for muon channel\\
			& $\pt^{\gamma}>20\GeV$, $\abs{\eta^{\gamma}}<1.444$ or $1.566<\abs{\eta^{\gamma}}<2.500$ \\
			& $\pt^{\mathrm{j1,j2}}>30\GeV$, $\abs{\eta^{\mathrm{j1,j2}}}<4.7$\\
			& $70<\mll<110\GeV$, $\mzg>100\GeV$\\
			& $\Delta R_{\mathrm{jj}}$, $\Delta R_{\mathrm{j}\gamma}$, $\Delta R_{\mathrm{j}\ell}>0.5$, $\Delta R_{\ell\gamma}>0.7$\\[\cmsTabSkip]
Control region & $150<\mjj<400\GeV$, \\
		& Common selection\\[\cmsTabSkip]
EW signal region	&$\mjj>500\GeV$, $\etajj>2.5$,\\
			& $\eta^*<2.4$, $\dphizgjj>1.9$,\\
			& Common selection\\[\cmsTabSkip]
Fiducial region		& $\mjj>500\GeV$, $\etajj>2.5$,\\
			& Common selection, without requirement on $\mzg$\\[\cmsTabSkip]
aQGC search region	& $\mjj>500\GeV$, $\etajj>2.5$,\\
			& $\pt^{\gamma}>100\GeV$, \\
			& Common selection, without requirement on $\mzg$\\
\hline
\end{tabular}
\label{tab:selections}
\end{table}

\section{Background estimation}
\label{sec:bkg_estimation}
The dominant source of background to the EW signal stems from QCD-induced $\PZ\gamma$jj production, such as the Feynman diagram in Fig.~\ref{fig:za_feynman}(lower right). The estimation of this background comes from simulation, and a simultaneous fit to the control and signal regions is used to constrain the uncertainties affecting its normalization. The uncertainties in the normalization of the QCD-induced $\PZ\gamma$jj are significantly smaller after this fit.

A background from events in which the selected photon is not prompt arises mainly from {\PZ}+jets production. This background is estimated by applying extrapolation factors to events in a nonprompt photon control sample in data enriched in {\PZ}+jets events that corresponds to each region defined in Table~\ref{tab:selections} through just a change in the photon selections. Instead of requiring the photon to pass the identification selection of medium working point, the photon is required to fail that but pass the more relaxed identification selection ~\cite{wgzg,Khachatryan:2017jub}. The nonprompt extrapolation factors are measured in data in a region similar to our common selection with the jet requirements removed. They are measured as a function of photon \pt, photon $\eta$, and lepton flavor; the typical variation ranges from 0.1 to 0.5. The numerator in the extrapolation factor is based on a template fit to the distribution in photon $\sigma_{\mathrm{i}\eta\mathrm{i}\eta}$ in data, through which the prompt and nonprompt photon contribution can be easily distinguished from each other. The variable $\sigma_{\mathrm{i}\eta\mathrm{i}\eta}$ quantifies the width of the photon electromagnetic shower in $\eta$, which is narrow for prompt and broad for nonprompt photons. The prompt template is obtained from simulated $\PZ\gamma$ events and the nonprompt template is obtained from a sideband of charged hadron isolation variable of photon in data. The denominator of the extrapolation factor is simply the number of events in the nonprompt photon control sample, since the contamination of the denominator by prompt photon events is negligible.

Other backgrounds estimated from simulation include single top quark events in the $s$- and $t$-channels that are normalized to their respective NLO cross sections; associated single top quark and {\PW} boson production normalized to its next-to-next-to-leading order (NNLO) cross section~\cite{STW_1}; {\PW}{\PW} production normalized to its NNLO cross section; {\PW}{\PZ}, {\PZ}{\PZ} and QCD-induced $\PW\gamma$jj production normalized to their NLO cross sections; and $\ttbar\gamma$ production normalized to its NLO cross section. All of these processes are also normalized to the integrated luminosity of the data.

\begin{figure}[ht]
   \centering
      \includegraphics[width=0.48\textwidth]{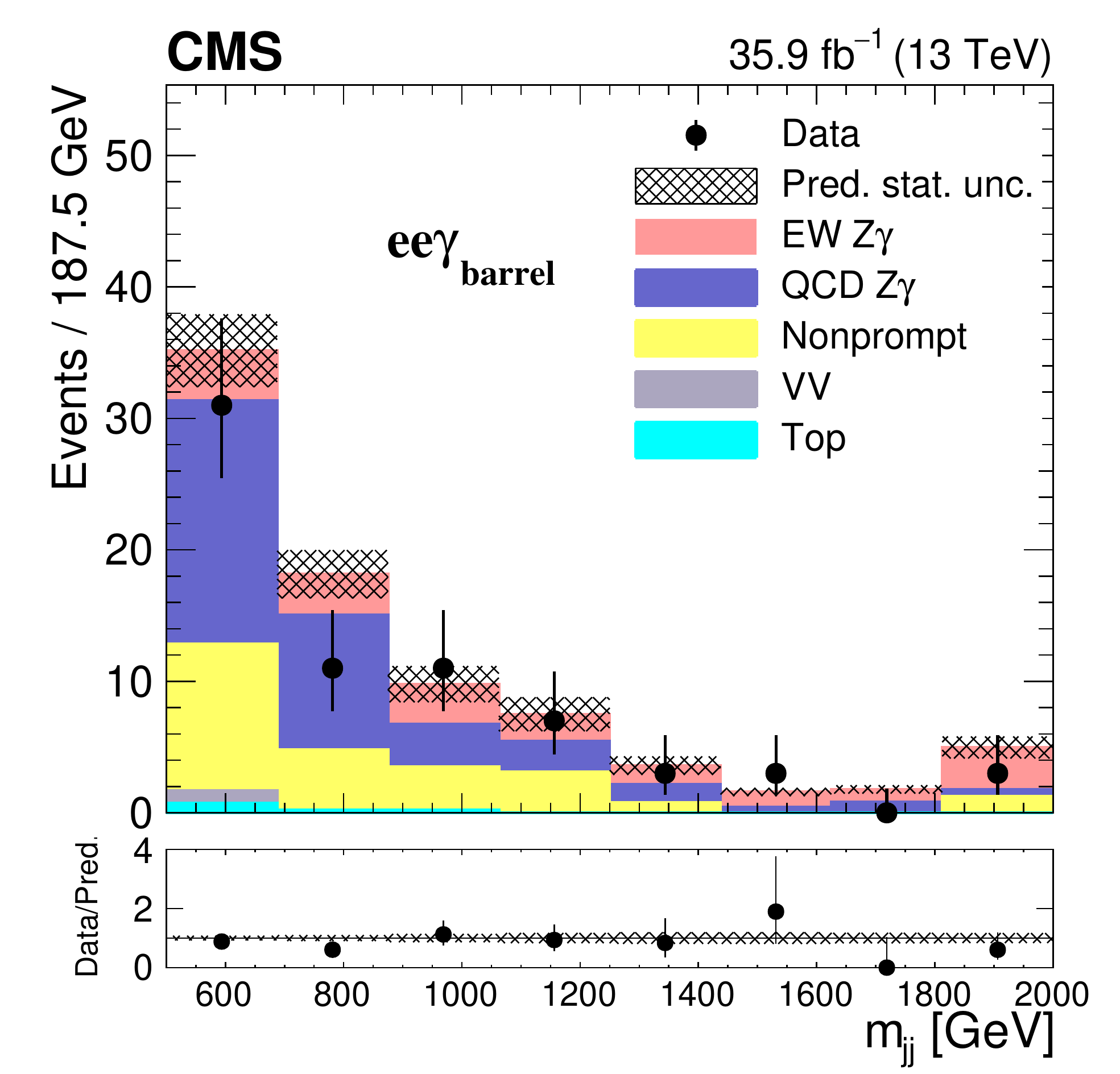}
      \includegraphics[width=0.48\textwidth]{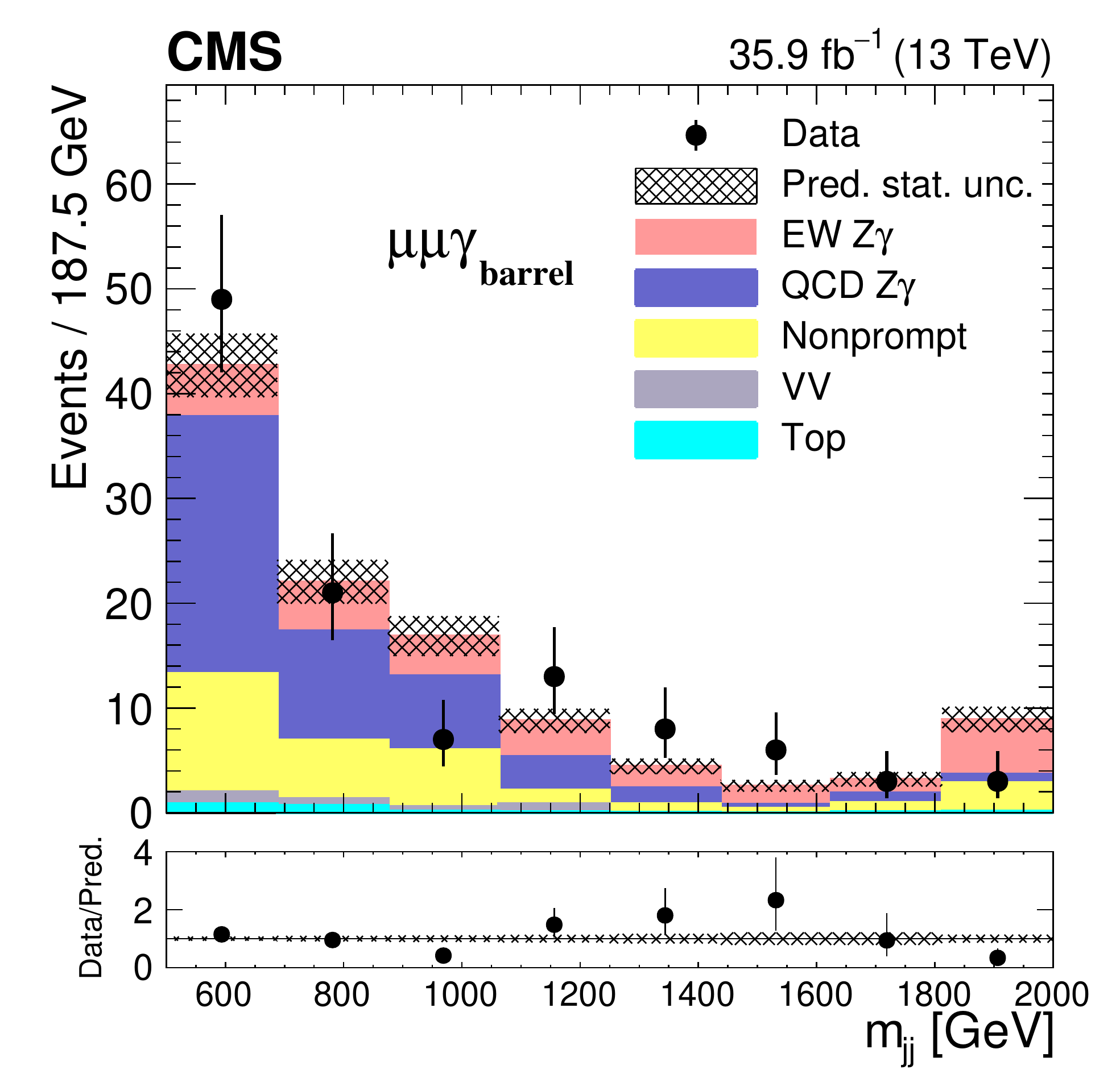}
      \caption{The pre-fit $\mjj$ distributions for the dilepton + $\gbarrel$ events are shown on the left for the dielectron and on the right for the dimuon categories. The data are compared to the sum of the signal and the background contribution. The black points with error bars represent the data and their uncertainties, while the hatched bands represent the statistical uncertainty on the combined signal and background expectations. The last bin includes overflow events. The bottom plots show the ratio of the data to the expectation.}
      \label{fig:mjj_distri_b}
\end{figure}

\begin{figure}[ht]
   \centering
      \includegraphics[width=0.48\textwidth]{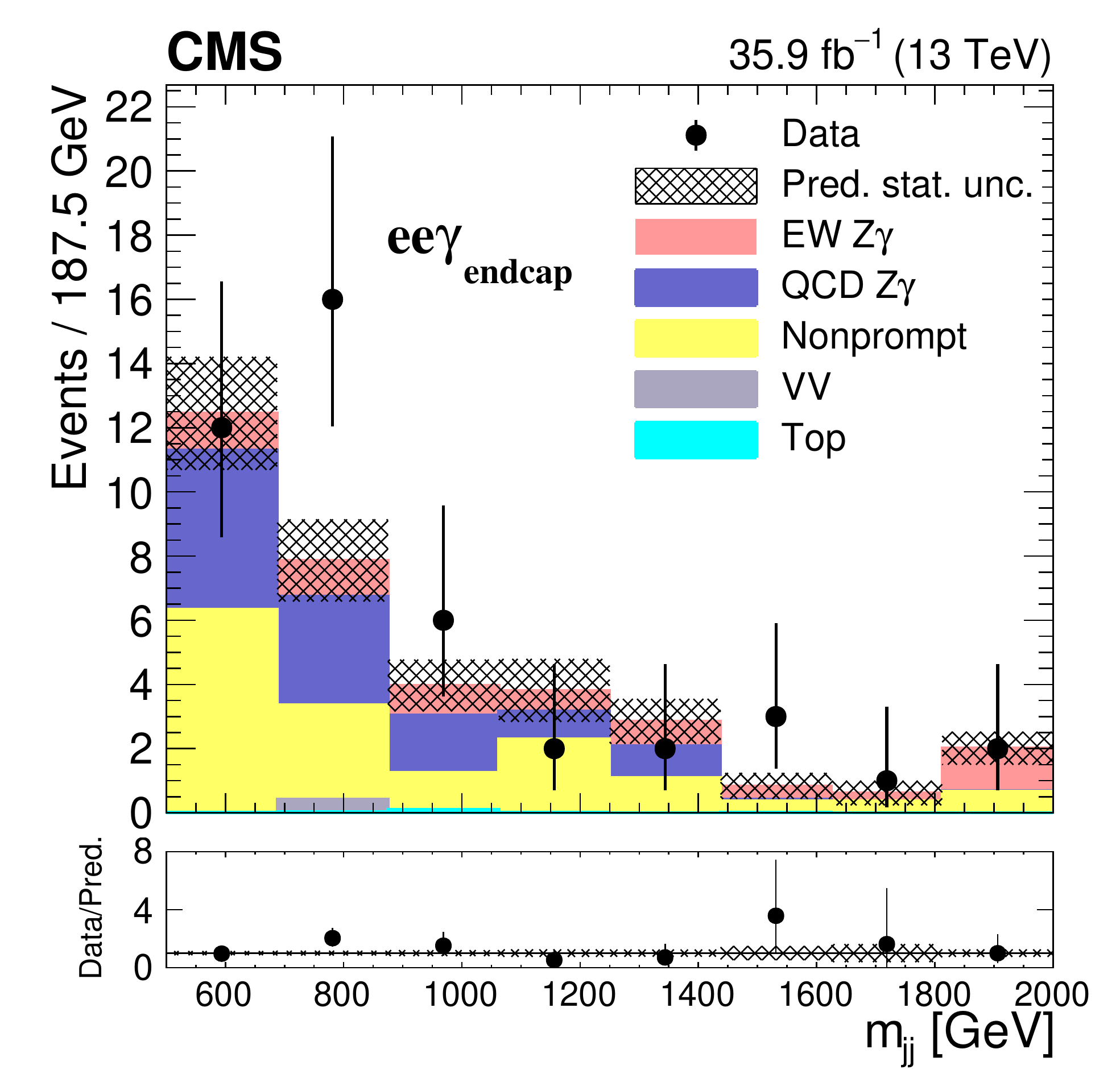}
      \includegraphics[width=0.48\textwidth]{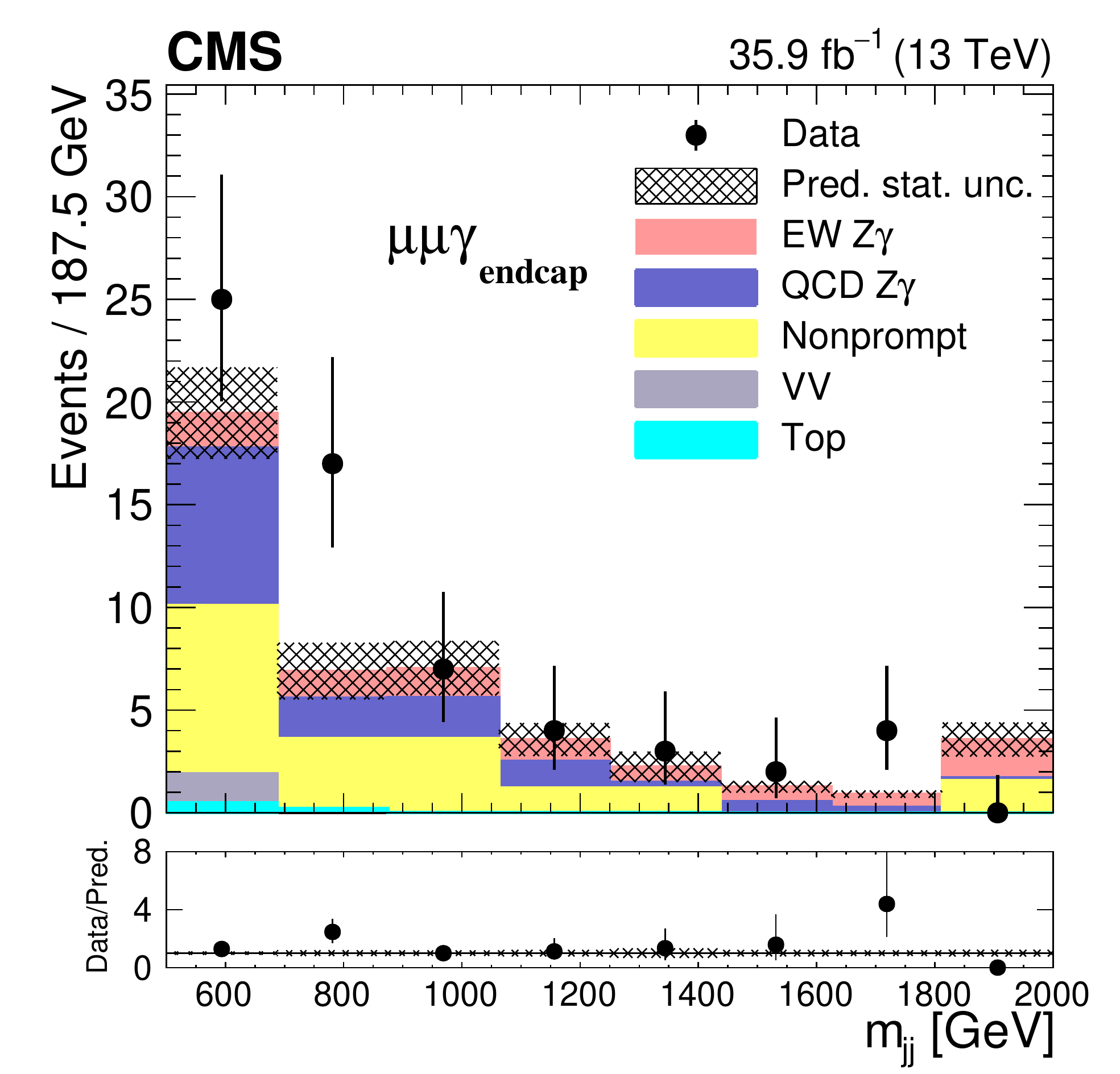}
      \caption{The pre-fit $\mjj$ distributions for the dilepton + $\gendcap$ events are shown on the left for the dielectron and on the right for the dimuon categories. The data are compared to the sum of the signal and the background contribution. The black points with error bars represent the data and their uncertainties, while the hatched bands represent the statistical uncertainty on the combined signal and background expectations. The last bin includes overflow events. The bottom plots show the ratio of the data to the expectation.}
      \label{fig:mjj_distri_e}
\end{figure}

After imposing the EW signal region selection, the pre-fit (\ie before the simultaneous fit) $\mjj$ distributions for the dilepton + $\gbarrel$ and the dilepton + $\gendcap$ categories described in Section~\ref{sec:event_selection} are shown respectively in Figs.~\ref{fig:mjj_distri_b} and ~\ref{fig:mjj_distri_e}. The agreement between data and the combined expectation for signal and backgrounds is reasonable.

\section{Systematic uncertainties}
\label{sec:uncertainties}

Systematic uncertainties that affect the measurements arise from experimental issues, such as detector effects and the methods used to compute higher-level quantities, \eg, efficiencies, and variations in theoretical inputs such as the choice of the renormalization and factorization scale and the choice of the PDFs. Each systematic uncertainty is quantified by evaluating its effect on the yield and distribution of relevant kinematic variables in the signal and background categories. The log-normal distribution is used to model the dependence on systematic uncertainties.

The systematic uncertainties in the trigger, lepton reconstruction, and selection efficiencies are measured using the tag-and-probe technique and are 2--3\%. The uncertainties in jet energy scale (JES) and jet energy resolution (JER) are calculated from simulated events by rescaling and spreading the jet \pt, and propagating the bin-by-bin effects in the variables. The uncertainties from JES and JER vary in the respective ranges of 1--49 and 1--26\%. An uncertainty of 2.5\% in the integrated luminosity~\cite{LUM-17-001} is estimated from simulation. The statistical uncertainties from the size of the number of simulated events as well as the size of data samples used in our background and signal are corrected assuming Poisson distributions, and calculated bin-by-bin. The uncertainties related to the number of simulated events, or to the limited number of events in the data control sample, are respectively 5--46\% for the EW $\PZ\gamma$jj signal, 10--50\% for the QCD-induced $\PZ\gamma$jj background, and 20--100\% for the nonprompt photon background where the uncertainty value increases with increasing $\mjj$ and $\etajj$, and are uncorrelated across different processes and bins of any single distribution. The uncertainties from the correction factors caused by the ECAL mistiming vary by 1--4\%, and are applied to all the simulated events and treated as being correlated across different processes and bins.

An overall uncertainty in the nonprompt photon background is estimated through the quadratic sum of systematic uncertainties from several sources. The uncertainty from the choice of isolation variable use in the sideband is estimated through the nonprompt photon fraction for alternative choices of isolation variable sideband~\cite{Khachatryan:2017jub}. An uncertainty on closure is defined by fits performed to the nonprompt photon fraction in simulated events and comparing the fit results with the known fractions. The closure uncertainty in the region of the endcap detector is larger than in the barrel, and becomes greater with increasing photon \pt. This uncertainty provides the dominant part of the systematic component from sources of nonprompt photons. The overall uncertainty in the nonprompt photon background is in the range of 9--37\%.

However, theoretical uncertainties have largest impact on the measurement. The scale uncertainty is estimated through simultaneous changes in the $\mu_\mathrm{R}$ and $\mu_\mathrm{F}$ scales up and down by a factor of two relative to their nominal value in each event, under the condition that 1/2 $\leq$ $\mu_\mathrm{R}$/$\mu_\mathrm{F}$ $\leq$ 2. The maximal difference with respect to the nominal value is taken as the measure of uncertainty. The uncertainties in the PDFs are estimated by combining the expectations from all of the contribution in the NNPDF3.0 set of PDFs, according to the procedure described in Ref.~\cite{Butterworth:2015oua}. For the signal, the scale uncertainty is within the range of 2--14\% and the PDF uncertainty within range 3--11\% that increases with increasing $\mjj$ and $\etajj$. The scale uncertainty in QCD-induced $\PZ\gamma$jj events, which has a large impact on the measurement, varies in the range of 5--25\%. It is constrained in the simultaneous fit to the signal in the low-$\mjj$ control region. The PDF uncertainty in the QCD-induced $\PZ\gamma$jj events is in the range of 1--3\%.

The interference term between the EW and QCD-induced processes at order $\alpha^4\alpS$ at the tree level, is estimated at the particle level using \MGvATNLO. The interference contribution is defined as the difference between the cross section for inclusive $\PZ\gamma$jj production, which contains the interference term, and the sum of the cross sections for pure EW $\PZ\gamma$jj and QCD-induced $\PZ\gamma$jj. It is positive, and the ratio of the interference to EW $\PZ\gamma$jj production that decreases with increasing $\mjj$ is in the range of 4--8\%, which is consistent with the range obtained from a pure interference term directly generated using \MGvATNLO.

All the above systematic uncertainties are applied to both the measured significance of the signal and to the search for aQGC. They are also propagated to the uncertainty in the measured fiducial cross section, with the exception of the theoretical uncertainties associated with the signal cross section. All systematic uncertainties except those arising from trigger and lepton identification efficiencies are assumed to be correlated between the electron and muon channels. Various sources of systematic uncertainties and their effect on the event yields in the process are summarized in Table~\ref{tab:nuisance}.

\begin{table}[htb]
\centering
\topcaption{The pre-fit systematic uncertainties in the measurement of the extracted signal. They are for signal or background (bkg) if the source is specified, or for both if the source is not specified.}
\begin{tabular}{lc}
\hline
Source of systematic uncertainty    & Relative uncertainty [\%]\\
\hline
Scales in QCD-induced $\PZ\gamma$jj bkg      & 5--25\\
Scales in EW $\PZ\gamma$ signal        & 2--14\\
Interference           & 4--8 \\
JES            & 1--49\\
JER             & 1--26 \\
Nonprompt photon bkg    &9--37       \\
Integrated luminosity & 2.5\\
L1 mistiming correction & 1--4\\
Photon identification  & 3\\
Pileup modeling		& 1\\
Trigger and selection efficiency& 2--3\\
\hline
\end{tabular}
\label{tab:nuisance}
\end{table}

\section{Results}
\subsection{Measurement of the signal significance}
\label{sec7}
The post-fit (\ie after the simultaneous fit) simulated signal and background yields as well as the observed data yields in the EW signal region are listed in Table~\ref{tab:yields}.

\begin{figure}[ht!]
   \centering
      \includegraphics[width=0.48\textwidth]{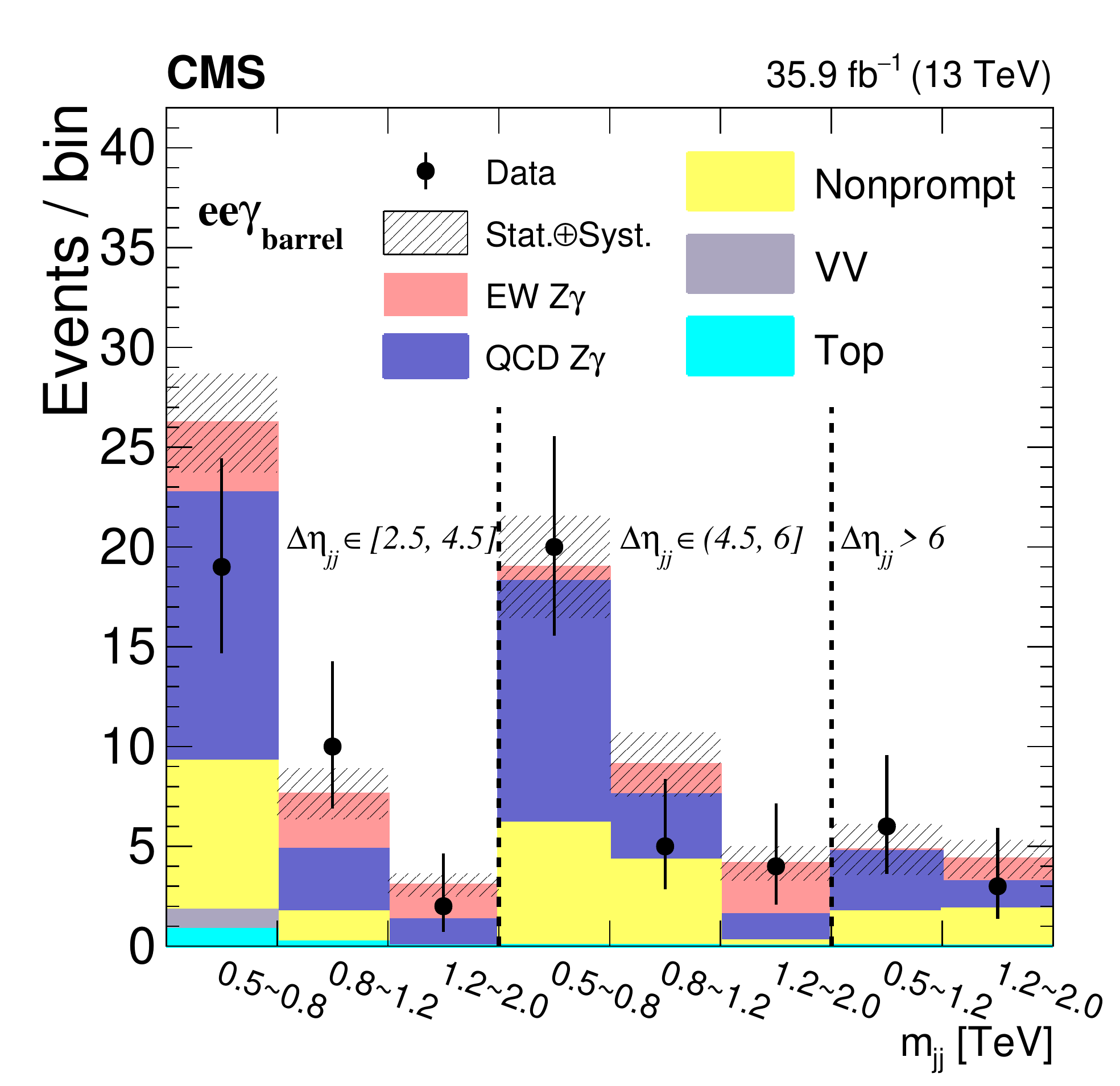}
      \includegraphics[width=0.48\textwidth]{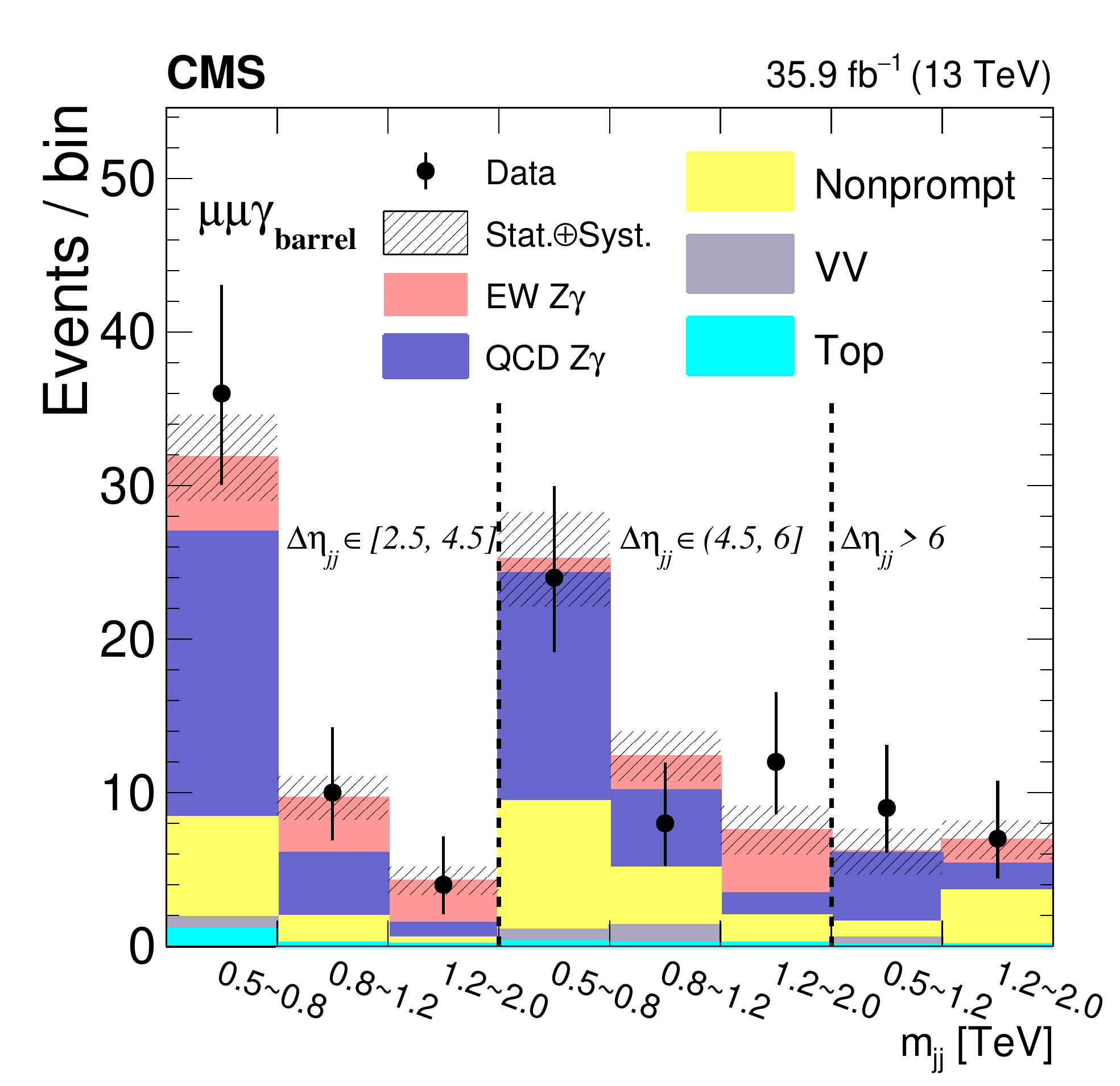}
      \caption{The post-fit 2D distributions of the dielectron (left) and dimuon (right) + $\gbarrel$ categories as a function of $\mjj$ in bins of $\etajj$. The horizontal axis is split into bins of $\etajj$ of [2.5, 4.5], (4.5, 6.0], and $>6.0$. The data are compared to the signal and background predictions in the signal region. The black points with error bars represent the data and statistical uncertainties of data, the hatched bands represent the full uncertainties of the predictions.}
      \label{fig:postfit_b}
\end{figure}

\begin{figure}[ht!]
   \centering
      \includegraphics[width=0.48\textwidth]{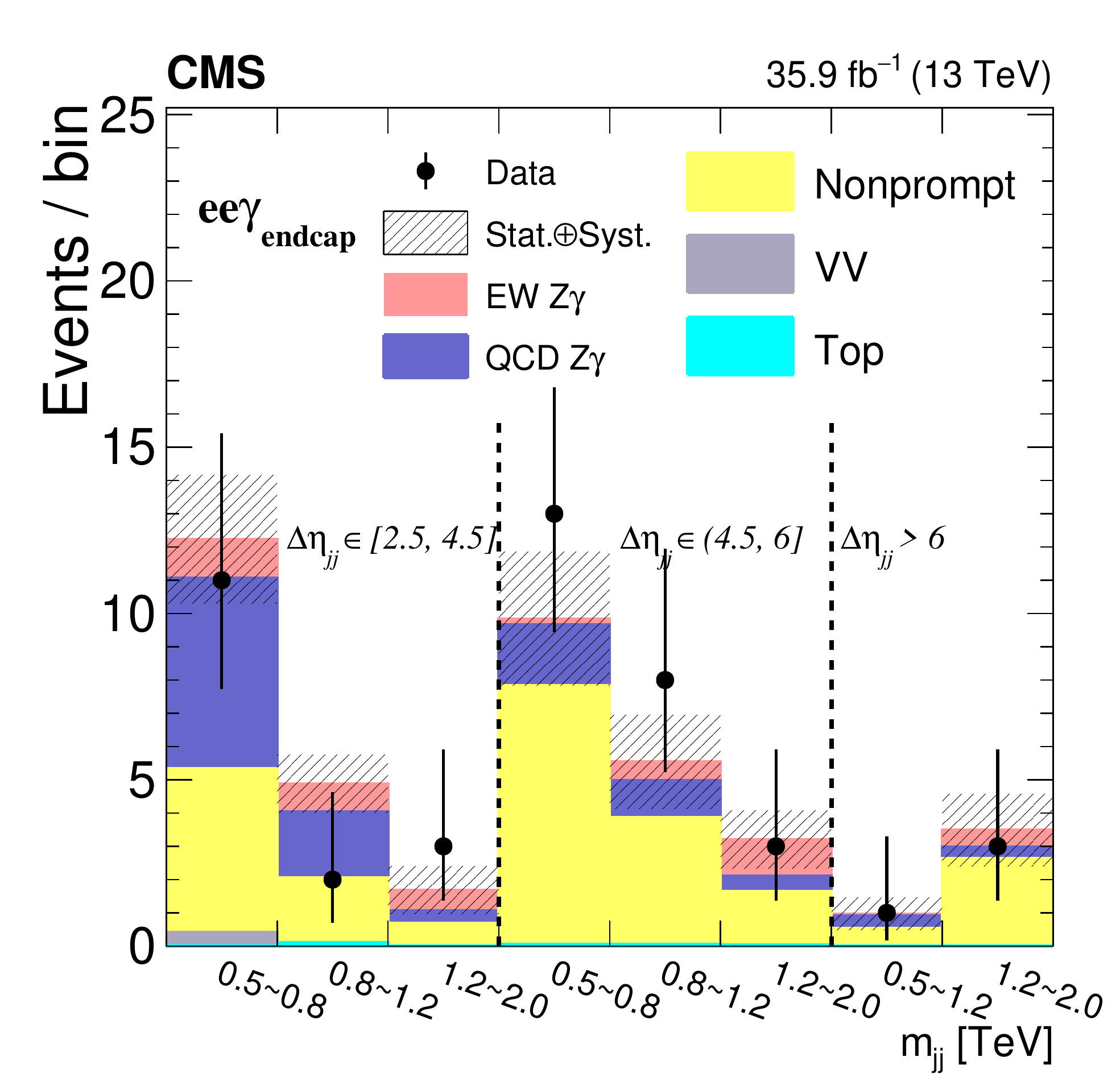}
      \includegraphics[width=0.48\textwidth]{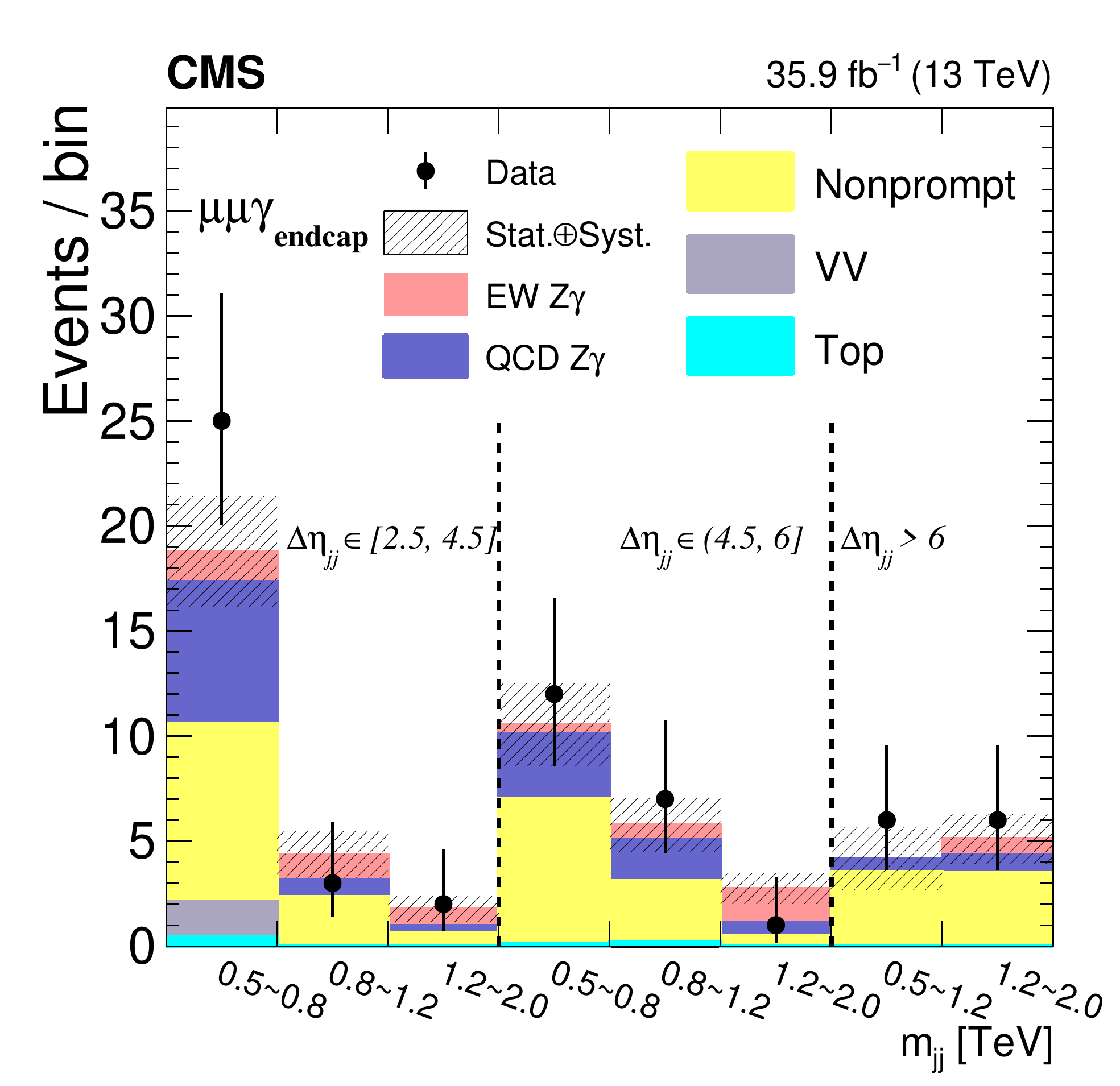}
      \caption{The post-fit 2D distributions of the dielectron (left) and dimuon (right) + $\gendcap$ categories as a function of $\mjj$ in bins of $\etajj$. The horizontal axis is split into bins of $\etajj$ of [2.5, 4.5], (4.5, 6.0], and $>6.0$. The data are compared to the signal and background predictions in the signal region. The black points with error bars represent the data and statistical uncertainties of data, the hatched bands represent the full uncertainties of the predictions.}
      \label{fig:postfit_bb}
\end{figure}

To quantify the significance of the measured EW $\PZ\gamma$ signal, a statistical analysis of the event yields is performed in a two-dimensional (2D) $\mjj$ and $\etajj$ grid. There are 4 categories within the signal region that correspond to the choice between barrel and endcap-detector photons and between electron and muon final states. For each bin in $\mjj$ and $\etajj$, we construct a Poisson function in the number of observed events. The likelihood is the product of the Poisson distributions for the bin contents and log-normal distributions for the uninteresting constraints in ``nuisance'' parameter. All background contributions are allowed to vary within their associated uncertainties. A $p$-value that represents the probability to obtain the data given a background-only hypothesis is computed using a profile likelihood-ratio test statistic~\cite{Junk,Read,CLs}. The $p$-value is then converted to a significance based on the area in the ``tail'' of a normal distribution. The post-fit 2D distributions are shown in Figs.~\ref{fig:postfit_b} and ~\ref{fig:postfit_bb}. The binning in $\mjj$ and $\etajj$ is optimized for best signal significance. The observed and expected significance for the signal in the data is 3.9 and 5.2 standard deviations with the data set collected in 2016. The main contributions to the significance are from bins with an excess of signal relative to background events, \ie, high $\mjj$ bins in each channel. The data in the dimuon + $\gbarrel$ and dielectron + $\gendcap$ channels are in good agreement with the expectations in these three bins, while the data are below the expectations in the other two channels. The downward fluctuations of the data in the dimuon + $\gendcap$ and dielectron + $\gbarrel$ channels result in the difference between the observed and expected significance. The total uncertainty on the measurement is dominated by the statistical uncertainty in the data. After combining this analysis with the results obtained at 8\TeV \cite{Khachatryan:2017jub} using a simultaneous fit, the observed and expected significance becomes, respectively, 4.7 and 5.5 standard deviations. In the combination of the 13\TeV and 8\TeV results, the theoretical uncertainties are treated as correlated because they affect the cross section of the sample and the calculation of the experimental acceptance in the same way, independently of the data-taking period; the experimental uncertainties in the efficiencies of the triggers, object reconstruction and identification are determined independently for each data sample and are uncorrelated.

\begin{table}[htb]
\centering
\topcaption{Post-fit signal and background yields and observed event counts in data after the final selection in the search for EW signal. The $\gbarrel$ and $\gendcap$ represent photons in the barrel and endcap-detector region, respectively. ``Other bkgs.'' represents the contribution of diboson, top and $\PW\gamma$ process. The uncertainties are the quadratic sum of statistical and systematic uncertainties.}
\begin{tabular}{lcccc}
\hline
Processes      & $\Pe\Pe\gamma_\mathrm{barrel}$ & $\Pe\Pe\gamma_\mathrm{endcap}$  &$\Pgm\Pgm\gamma_\mathrm{barrel}$  & $\Pgm\Pgm\gamma_\mathrm{endcap}$\\
\hline
QCD-induced $\PZ\gamma$jj bkg. & $39.0\pm 3.0$ &  $12.2\pm 1.4$ & $51.1\pm 3.5$ & $14.9\pm 1.5$\\
Nonprompt photon bkg. & $23.2\pm 3.0$ & $23.9\pm 3.3$ & $27.1\pm 3.2$ & $28.9\pm 3.8$\\
Other bkgs. & $2.2\pm 1.0$ & $0.7\pm 0.5$ & $5.4\pm 1.3$ & $2.5\pm 1.0$\\
Total bkgs. & $64.4\pm 4.4$ & $36.8\pm 3.6$ & $83.6\pm 5.0$ & $46.3\pm 4.2$\\
EW $\PZ\gamma$jj signal & $14.0\pm 1.6$ & $5.0\pm 0.6$ & $20.2\pm 2.3$ & $7.0\pm 0.8$\\[\cmsTabSkip]
EW signal + total bkgs. & $78.4\pm 4.7$ & $41.8\pm 3.7$ & $103.8\pm 5.5$ & $53.3\pm 4.3$\\[\cmsTabSkip]
Data			& 69 & 44  & 110 & 62 \\
\hline
\end{tabular}
\label{tab:yields}
\end{table}


\subsection{Fiducial cross section}
A fiducial cross section is extracted using the same $\mjj$--$\etajj$ binnings as used in the calculation of the significance, and through the same simultaneous fit used in the control region. The fiducial region is defined in Table~\ref{tab:selections}. We define the cross section as
\begin{linenomath}
\begin{equation*}\label{sigma1}
\sigma^{\text{fid}} = \sigma_{\mathrm{g}} \hat{\mu} \mathrm{a}_{\mathrm{gf}},
\end{equation*}
\end{linenomath}
where $\sigma_\mathrm{g}$ is the cross section for the generated signal events, $\hat{\mu}$ is the signal strength parameter, and $\mathrm{a}_\mathrm{gf}$ is the acceptance for the events generated in the fiducial region and evaluated through simulation. The fiducial cross section for the EW $\PZ\gamma$ signal obtained from \MGvATNLO at LO accuracy is $4.97\pm 0.25 \mathrm{ (scale)}\pm 0.14 \mathrm{ (PDF)\unit{fb}}$. The best fit value for the EW $\PZ\gamma$ signal strength is $0.65\pm 0.24$ and the measured fiducial cross section is
\begin{linenomath}
\begin{equation*}\label{method1}
\sigma^{\text{fid}}_{\mathrm{EW}} = 3.2\pm 0.2\lum\pm 1.1\stat\pm 0.6\syst\unit{fb} = 3.2\pm 1.2\unit{fb}.
\end{equation*}
\end{linenomath}
A combined $\PZ\gamma$jj cross section is measured in the same fiducial region using the same procedure, except that the control region is excluded. The combined $\PZ\gamma$jj cross section is defined as
\begin{linenomath}
\begin{equation*}\label{fidall}
\sigma^{\text{fid}} = \hat{\mu} \{ \sigma^{\mathrm{EW}}_{\mathrm{g}} \mathrm{a}^{\mathrm{EW}}_{\mathrm{gf}} + \sigma^{\mathrm{QCD}}_{\mathrm{g}} \mathrm{a}^{\mathrm{QCD}}_{\mathrm{gf}} \}.
\end{equation*}
\end{linenomath}
The fiducial cross section for all QCD-induced $\PZ\gamma$jj events expected from \MGvATNLO at NLO accuracy is $10.7\pm 1.7\,\text{(scale)}\pm 0.2\,\mathrm{(PDF)\unit{fb}}$. The expected fiducial cross section for the combined QCD and EW $\PZ\gamma$jj production is $15.7\pm 1.7\,\text{(scale)}\pm 0.2 \mathrm{(PDF)\unit{fb}}$. The best fit value for the combined $\PZ\gamma$jj signal strength is $0.91\pm 0.19$, and the measured cross section is
\begin{linenomath}
\begin{equation*}\label{fidall2}
\sigma^{\text{fid}}_{\mathrm{EW+QCD}} = 14.3\pm 0.4\lum\pm 1.1\stat\pm 2.7\syst\unit{fb} = 14.3\pm 3.0\unit{fb}.
\end{equation*}
\end{linenomath}

\subsection{Limits on anomalous quartic gauge couplings}
The effects of BSM physics can be modeled in a generic way through a collection of linearly independent higher dimensional operators in effective field theory~\cite{paper_aqgc}. Reference~\cite{Eboli:2006wa} proposes nine independent charge-conjugate and parity-conserving dimension-eight effective operators by assuming the SU(2)$\times$U(1) symmetry of the EW gauge field, including a Higgs doublet to incorporate the presence of an SM Higgs boson. A contribution from aQGCs would enhance the production of events with large $\PZ\gamma$ mass. The operators affecting the $\PZ\gamma$jj channel can be divided into those containing an SU(2) field strength, the U(1) field strength, the covariant derivative of the Higgs doublet, ${\cal L}_{\mathrm{M,0}} - {\cal L}_{\mathrm{M,7}}$, and those containing only the two field strengths, ${\cal L}_{\mathrm{T,0}} - {\cal L}_{\mathrm{T,9}}$. The coefficient of the operator ${\cal L}_{\mathrm{X,Y}}$ is denoted by $F_{\mathrm{X,Y}}/\Lambda^4$, where $\Lambda$ is the unknown scale of BSM physics.

\begin{figure}[h]
   \centering
      \includegraphics[width=0.9\textwidth]{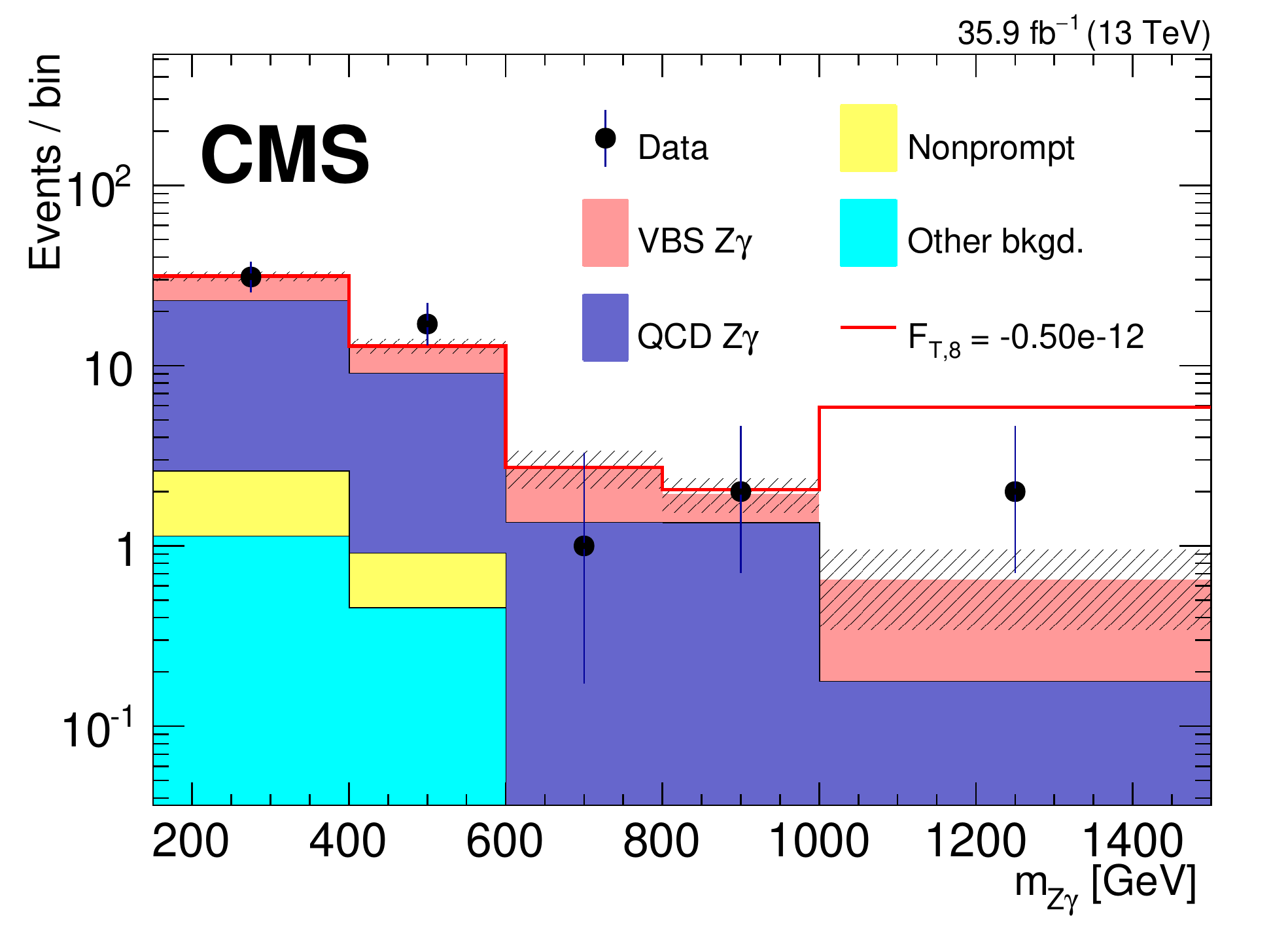}
      \caption{The $\mzg$ distribution of events satisfying the aQGC region selection, which is used to set constraints on the anomalous coupling parameters. The red line represents a nonzero $F_{\mathrm{T,8}}$ setting, which would significantly enhance the yields at high $\mzg$. The bins of $\mzg$ are [100, 400, 600, 800, 1000, 1500]\GeV, where the last bin includes overflow. The hatched bands represent the statistical uncertainties in the predictions.}
      \label{fig:aqgc_yields}
\end{figure}

A simulation is performed that includes the effects of the aQGCs in addition to the SM EW $\PZ\gamma$ process, as well as any interference between the two. We use the $\mzg$ distribution to extract limits on aQGC parameters. To obtain a continuous prediction for the signal as a function of the anomalous coupling, a quadratic fit is performed to the SM+aQGC yield as a function of $\mzg$ bin in the aQGC region defined in Section~\ref{sec:event_selection}. From Fig.~\ref{fig:aqgc_yields}, no statistically significant excess of events relative to the SM prediction.
The following profile likelihood test statistic is used in the aQGC limit setting procedure:
\begin{linenomath}
\begin{equation*}\label{likelihoodformula}
t_{\alpha_\text{test}} = -2 \log \frac{{\cal L}(\alpha_\text{test},{\hat{\hat{\boldsymbol{\theta}}}})}{\mathcal{L}(\hat{\alpha},\hat{\boldsymbol{\theta}})}.
\end{equation*}
\end{linenomath}
The likelihood function is the product of Poisson distributions and a normal constraining term with nuisance parameters representing the sources of systematic uncertainties in any given bin. The final likelihood function is the product of the likelihood functions of the electron and muon channels. The main constraint on aQGCs parameter is from the last bin. The $\alpha_{\text{test}}$ represents the aQGC point being tested, and the symbol $\boldsymbol{\theta}$ represents a vector of nuisance parameters assumed to follow log-normal distributions. The parameter $\hat{\hat{\boldsymbol{\theta}}}$ corresponds to the maximum of the likelihood function at the point $\alpha_{\text{test}}$. The $\hat{\alpha}$ and $\hat{\boldsymbol{\theta}}$ parameters correspond to the global maximum of the likelihood function.

\begin{figure*}[h]
   \centering
      \includegraphics[width=0.48\textwidth]{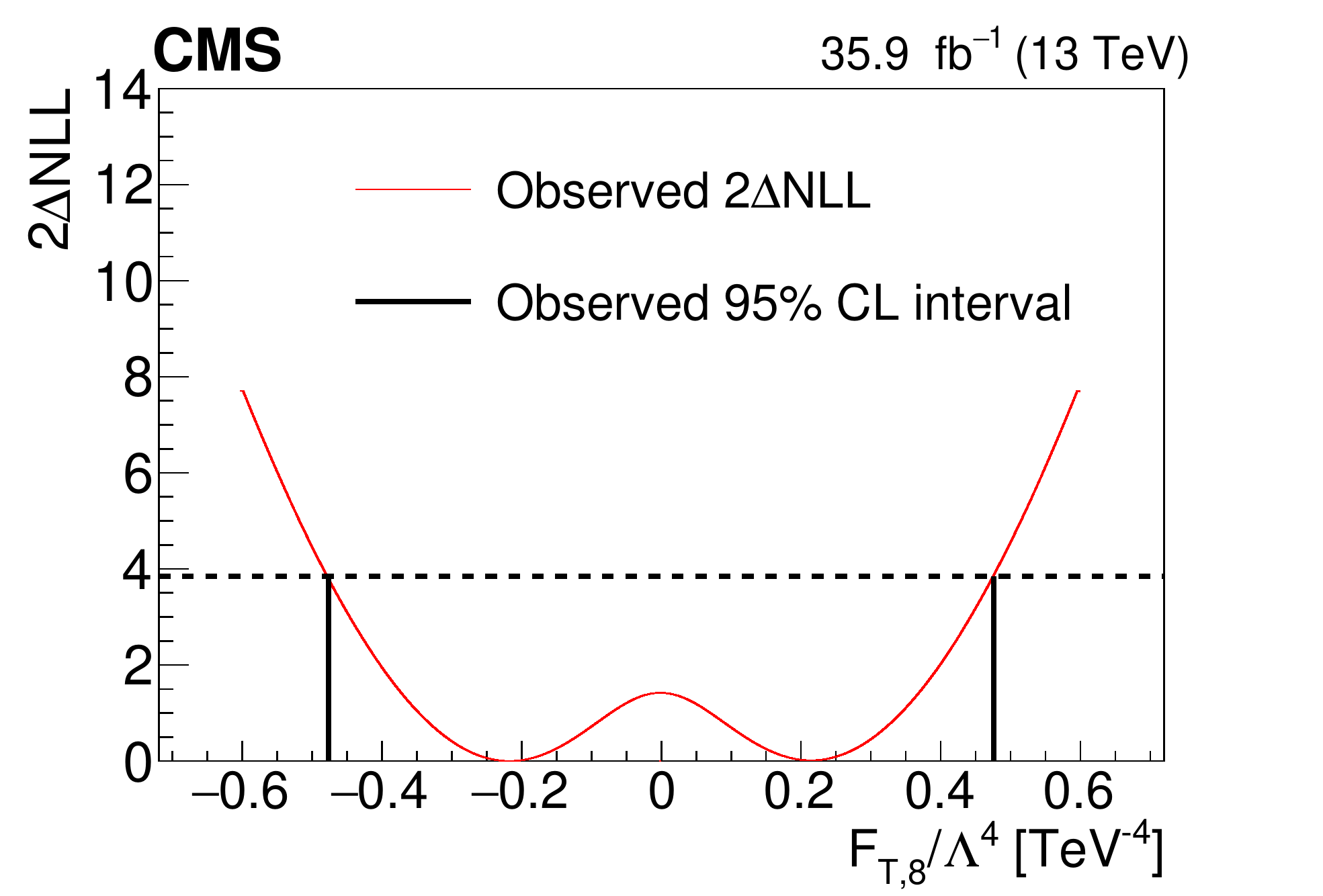}
      \includegraphics[width=0.48\textwidth]{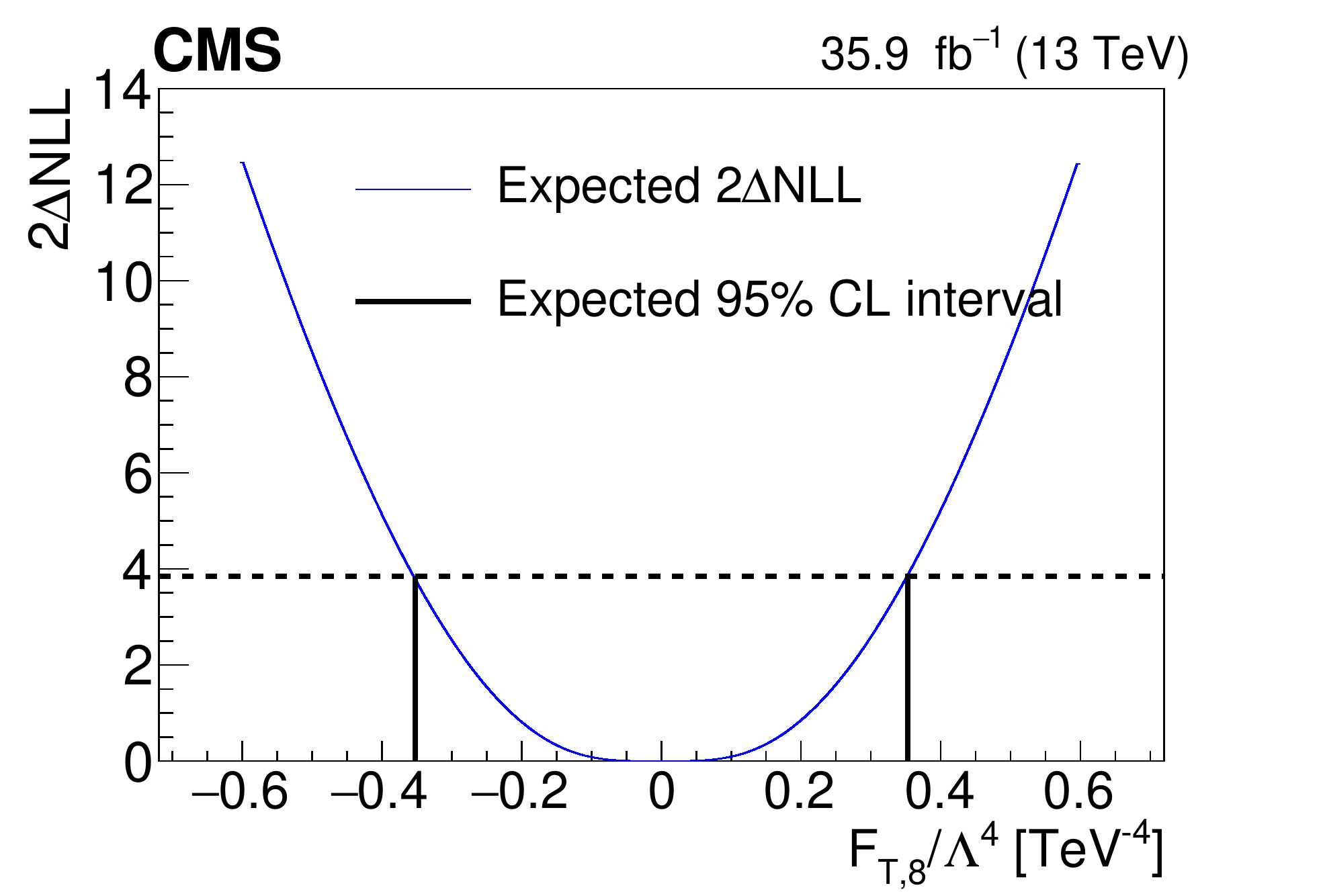}
      \caption{Observed (left) and expected (right) 95\% \CL intervals on the aQGC parameter $F_{\mathrm{T,8}}$.}
      \label{fig:aqgc_limit}
\end{figure*}

This test statistic is assumed to follow a $\chi^2$ distribution~\cite{1943Wald:wilks1938}. It is therefore possible to extract the limits immediately from the difference in the log-likelihood function $\Delta\mathrm{NLL} = t_{\alpha_{\text{test}}}/2$~\cite{Khachatryan:2014jba}. The 95\% confidence level (\CL) limit on a one dimensional aQGC parameter corresponds to $2\Delta\mathrm{NLL} = 3.84$. Figure~\ref{fig:aqgc_limit} shows the likelihood scan of parameter $F_{\mathrm{T,8}}$ in the calculation of the observed and expected limits.
The observed and expected 95\% \CL limits for the coefficients, shown in Table~\ref{tab:VBS_aQGC}, are obtained by varying the coefficients of one nonzero operator coefficient at a time. The observed limits are less stringent than those expected because of an excess of events at large $\mzg$, where you would expect aQGC signal, at approximately one standard deviation level. The unitarity bound is defined as the scattering energy at which the aQGC coupling strength set equal to the observed limit would result in a scattering amplitude that violates unitarity. The value of the unitarity bound is determined using the \textsc{vbfnlo}~2.7.1 framework~\cite{VBFNLO}, taking into account the difference between \textsc{vbfnlo} and \MGvATNLO. These results provide the most stringent limits to date on the aQGC parameters $F_{\mathrm{T,8}}/\Lambda^4$ and $F_{\mathrm{T,9}}/\Lambda^4$.

\begin{table}[htb]
\centering
  \topcaption{95\% \CL exclusion limits in units of \TeV$^{-4}$; the unitarity bounds are also listed in units of \TeV.}
\scalebox{1.0}{
  \begin{tabular}{ccc}
  \hline
  Observed limits [\TeV$^{-4}$] & Expected limits [\TeV$^{-4}$] & Unitarity bound [\TeV]\\
  \hline
	     $-19.5<F_{\mathrm{M,0}}/\Lambda^{4}<20.3$  & $-15.0<F_{\mathrm{M,0}}/\Lambda^{4}<15.0$   & 1.0  \\
             $-40.5<F_{\mathrm{M,1}}/\Lambda^{4}<39.5$  & $-30.0<F_{\mathrm{M,1}}/\Lambda^{4}<29.9$   & 1.2  \\
             $-8.22<F_{\mathrm{M,2}}/\Lambda^{4}<8.10$  & $-6.09<F_{\mathrm{M,2}}/\Lambda^{4}<6.06$     & 1.3  \\
             $-17.7<F_{\mathrm{M,3}}/\Lambda^{4}<17.9$  & $-13.1<F_{\mathrm{M,3}}/\Lambda^{4}<13.2$    & 1.4  \\
             $-15.3<F_{\mathrm{M,4}}/\Lambda^{4}<15.8$  & $-11.7<F_{\mathrm{M,4}}/\Lambda^{4}<11.7$   & 1.4  \\
             $-25.1<F_{\mathrm{M,5}}/\Lambda^{4}<24.5$  & $-19.0<F_{\mathrm{M,5}}/\Lambda^{4}<18.1$  & 1.8  \\
             $-38.9<F_{\mathrm{M,6}}/\Lambda^{4}<40.6$  & $-29.9<F_{\mathrm{M,6}}/\Lambda^{4}<30.0$   & 1.0  \\
             $-60.3<F_{\mathrm{M,7}}/\Lambda^{4}<62.5$  & $-45.9<F_{\mathrm{M,7}}/\Lambda^{4}<46.1$    & 1.3  \\
             $-0.74<F_{\mathrm{T,0}}/\Lambda^{4}<0.69$  & $-0.56<F_{\mathrm{T,0}}/\Lambda^{4}<0.51$     & 1.4  \\
             $-0.98<F_{\mathrm{T,1}}/\Lambda^{4}<0.96$  & $-0.72<F_{\mathrm{T,1}}/\Lambda^{4}<0.72$     & 1.4  \\
             $-1.97<F_{\mathrm{T,2}}/\Lambda^{4}<1.86$  & $-1.47<F_{\mathrm{T,2}}/\Lambda^{4}<1.37$     & 1.4  \\
             $-0.70<F_{\mathrm{T,5}}/\Lambda^{4}<0.75$  & $-0.51<F_{\mathrm{T,5}}/\Lambda^{4}<0.57$     & 1.7  \\
             $-1.64<F_{\mathrm{T,6}}/\Lambda^{4}<1.68$  & $-1.23<F_{\mathrm{T,6}}/\Lambda^{4}<1.26$     & 1.6  \\
             $-2.59<F_{\mathrm{T,7}}/\Lambda^{4}<2.82$  & $-1.91<F_{\mathrm{T,7}}/\Lambda^{4}<2.12$     & 1.7  \\
             $-0.47<F_{\mathrm{T,8}}/\Lambda^{4}<0.47$  & $-0.36<F_{\mathrm{T,8}}/\Lambda^{4}<0.36$    & 1.5  \\
             $-1.27<F_{\mathrm{T,9}}/\Lambda^{4}<1.27$  & $-0.94<F_{\mathrm{T,9}}/\Lambda^{4}<0.94$    & 1.5  \\
  \hline
  \end{tabular}}
  \label{tab:VBS_aQGC}
\end{table}

\section{Summary}
A new measurement has been made of vector boson scattering in the $\PZ\gamma$jj channel. The data, collected in proton-proton collisions at $\sqrt{s}$ = 13\TeV in the CMS detector in 2016, correspond to an integrated luminosity of 35.9\fbinv. Events were selected by requiring two identified oppositely charged electrons or muons with invariant mass consistent with a {\PZ} boson, one identified photon, and two jets that have a large separation in pseudorapidity and a large dijet mass. The observed significance for a signal in the data is 3.9 standard deviations (s.d.), where a significance of 5.2 s.d. is expected based on the standard model. When this result is combined with previous CMS measurements at 8\TeV, the observed and expected significances become respectively 4.7 and 5.5 s.d. The fiducial cross section for electroweak $\PZ\gamma$jj production is $3.2\pm 1.2$\unit{fb} for the data at 13\TeV, and the fiducial cross section for the sum of sources from electroweak and from quantum chromodynamics is $14.3\pm 3.0$\unit{fb}.
Constraints placed on anomalous quartic gauge couplings in terms of dimension-eight operators in effective field theory are either competitive with or more stringent than those previously obtained.

\begin{acknowledgments}
We congratulate our colleagues in the CERN accelerator departments for the excellent performance of the LHC and thank the technical and administrative staffs at CERN and at other CMS institutes for their contributions to the success of the CMS effort. In addition, we gratefully acknowledge the computing centers and personnel of the Worldwide LHC Computing Grid for delivering so effectively the computing infrastructure essential to our analyses. Finally, we acknowledge the enduring support for the construction and operation of the LHC and the CMS detector provided by the following funding agencies: BMBWF and FWF (Austria); FNRS and FWO (Belgium); CNPq, CAPES, FAPERJ, FAPERGS, and FAPESP (Brazil); MES (Bulgaria); CERN; CAS, MoST, and NSFC (China); COLCIENCIAS (Colombia); MSES and CSF (Croatia); RPF (Cyprus); SENESCYT (Ecuador); MoER, ERC IUT, PUT and ERDF (Estonia); Academy of Finland, MEC, and HIP (Finland); CEA and CNRS/IN2P3 (France); BMBF, DFG, and HGF (Germany); GSRT (Greece); NKFIA (Hungary); DAE and DST (India); IPM (Iran); SFI (Ireland); INFN (Italy); MSIP and NRF (Republic of Korea); MES (Latvia); LAS (Lithuania); MOE and UM (Malaysia); BUAP, CINVESTAV, CONACYT, LNS, SEP, and UASLP-FAI (Mexico); MOS (Montenegro); MBIE (New Zealand); PAEC (Pakistan); MSHE and NSC (Poland); FCT (Portugal); JINR (Dubna); MON, RosAtom, RAS, RFBR, and NRC KI (Russia); MESTD (Serbia); SEIDI, CPAN, PCTI, and FEDER (Spain); MOSTR (Sri Lanka); Swiss Funding Agencies (Switzerland); MST (Taipei); ThEPCenter, IPST, STAR, and NSTDA (Thailand); TUBITAK and TAEK (Turkey); NASU (Ukraine); STFC (United Kingdom); DOE and NSF (USA).

\hyphenation{Rachada-pisek} Individuals have received support from the Marie-Curie program and the European Research Council and Horizon 2020 Grant, contract Nos.\ 675440, 752730, and 765710 (European Union); the Leventis Foundation; the A.P.\ Sloan Foundation; the Alexander von Humboldt Foundation; the Belgian Federal Science Policy Office; the Fonds pour la Formation \`a la Recherche dans l'Industrie et dans l'Agriculture (FRIA-Belgium); the Agentschap voor Innovatie door Wetenschap en Technologie (IWT-Belgium); the F.R.S.-FNRS and FWO (Belgium) under the ``Excellence of Science -- EOS" -- be.h project n.\ 30820817; the Beijing Municipal Science \& Technology Commission, No. Z191100007219010; the Ministry of Education, Youth and Sports (MEYS) of the Czech Republic; the Deutsche Forschungsgemeinschaft (DFG) under Germany's Excellence Strategy -- EXC 2121 ``Quantum Universe" -- 390833306; the Lend\"ulet (``Momentum") Program and the J\'anos Bolyai Research Scholarship of the Hungarian Academy of Sciences, the New National Excellence Program \'UNKP, the NKFIA research grants 123842, 123959, 124845, 124850, 125105, 128713, 128786, and 129058 (Hungary); the Council of Science and Industrial Research, India; the HOMING PLUS program of the Foundation for Polish Science, cofinanced from European Union, Regional Development Fund, the Mobility Plus program of the Ministry of Science and Higher Education, the National Science Center (Poland), contracts Harmonia 2014/14/M/ST2/00428, Opus 2014/13/B/ST2/02543, 2014/15/B/ST2/03998, and 2015/19/B/ST2/02861, Sonata-bis 2012/07/E/ST2/01406; the National Priorities Research Program by Qatar National Research Fund; the Ministry of Science and Education, grant no. 14.W03.31.0026 (Russia); the Tomsk Polytechnic University Competitiveness Enhancement Program and ``Nauka" Project FSWW-2020-0008 (Russia); the Programa Estatal de Fomento de la Investigaci{\'o}n Cient{\'i}fica y T{\'e}cnica de Excelencia Mar\'{\i}a de Maeztu, grant MDM-2015-0509 and the Programa Severo Ochoa del Principado de Asturias; the Thalis and Aristeia programs cofinanced by EU-ESF and the Greek NSRF; the Rachadapisek Sompot Fund for Postdoctoral Fellowship, Chulalongkorn University and the Chulalongkorn Academic into Its 2nd Century Project Advancement Project (Thailand); the Kavli Foundation; the Nvidia Corporation; the SuperMicro Corporation; the Welch Foundation, contract C-1845; and the Weston Havens Foundation (USA).
\end{acknowledgments}

\bibliography{auto_generated}
\cleardoublepage \appendix\section{The CMS Collaboration \label{app:collab}}\begin{sloppypar}\hyphenpenalty=5000\widowpenalty=500\clubpenalty=5000\vskip\cmsinstskip
\textbf{Yerevan Physics Institute, Yerevan, Armenia}\\*[0pt]
A.M.~Sirunyan$^{\textrm{\dag}}$, A.~Tumasyan
\vskip\cmsinstskip
\textbf{Institut f\"{u}r Hochenergiephysik, Wien, Austria}\\*[0pt]
W.~Adam, F.~Ambrogi, T.~Bergauer, J.~Brandstetter, M.~Dragicevic, J.~Er\"{o}, A.~Escalante~Del~Valle, M.~Flechl, R.~Fr\"{u}hwirth\cmsAuthorMark{1}, M.~Jeitler\cmsAuthorMark{1}, N.~Krammer, I.~Kr\"{a}tschmer, D.~Liko, T.~Madlener, I.~Mikulec, N.~Rad, J.~Schieck\cmsAuthorMark{1}, R.~Sch\"{o}fbeck, M.~Spanring, D.~Spitzbart, W.~Waltenberger, C.-E.~Wulz\cmsAuthorMark{1}, M.~Zarucki
% ...
\vskip\cmsinstskip
\dag: Deceased\\
1:  Also at Vienna University of Technology, Vienna, Austria\\
% ...
\end{sloppypar}
%%% END EDITABLE REGION %%%
\end{document}